\pdfoutput=1

\documentclass[11pt,a4paper]{article}

\usepackage{jheppub}
\usepackage{amsmath}
\usepackage{multicol,bbm}
\usepackage{enumerate}
\usepackage{bibentry}

\usepackage{graphicx}
\usepackage{dcolumn}
\usepackage{float}
\usepackage[many]{tcolorbox}
\usepackage{lipsum}
\usepackage{cancel}

\usepackage{shuffle}
\usepackage{graphbox}
\usepackage{environ}
\usepackage{physics}

\usepackage{amssymb}
\usepackage{amsthm}

\usepackage{bm}
\usepackage{multirow}
\usepackage{caption}
\usepackage{tabularx}
\usepackage{enumitem}

\usepackage{extarrows}
\usepackage{tikz}
\usepackage{verbatim} 

\newcommand{\inp}[2]{{(#1\cdot #2)}}

\graphicspath{{figures/}}

\begin{document}


\title{Integrating the full four-loop negative geometries and all-loop ladder-type negative geometries in ABJM theory}


\date{\today}

\author[a]{Zhenjie Li}

\affiliation[a]{SLAC National Accelerator Laboratory, Stanford University, Stanford, CA 94309, USA}
\emailAdd{zhenjiel@slac.stanford.edu}

\preprint{SLAC-PUB-17766}

\date{\today}

\abstract{The decomposition of the four-point ABJM amplituhedron into negative geometries produces compact integrands of logarithmic of amplitudes such that the infrared divergence only comes from the last loop integration, from which we can compute the cusp anomalous dimension of the ABJM theory. In this note, we integrate $L-1$ loop momenta of the $L$-loop negative geometries for all four-loop negative geometries and a special class of all-loop ladder-type negative geometries by a method based on Mellin transformation, and from these finite quantities we extract the corresponding contribution to the cusp anomalous dimension. We find that the infrared divergence of a box-type negative geometry at $L=4$ is weaker than other negative geometries, then only tree-type negative geometries contribute to the cusp anomalous dimension at $L=4$. For the all-loop ladder-type negative geometries, we prove and conjecture some recursive structures as integral equations in Mellin space and find that they cannot contribute zeta values like $\zeta_3,\zeta_5$ to the cusp anomalous dimension.}


\maketitle

\section{Introduction and review}\label{sec1}

The amplituhedron of planar ${\cal N}=4$ super Yang-Mills theory (SYM)~\cite{Arkani-Hamed:2013jha,Arkani-Hamed:2013kca, Arkani-Hamed:2017vfh} is a surprising geometric structure, where the canonical forms of these positive geometries~\cite{Arkani-Hamed:2017tmz} encode all-loop, all-multiplicity scattering amplitudes in the theory. This discovery has spurred the exploration of analogous positive geometries in various other theories and contexts (see, for example, \cite{Arkani-Hamed:2017fdk, Arkani-Hamed:2017mur, Arkani-Hamed:2019mrd, Arkani-Hamed:2019vag, Arkani-Hamed:2019plo, Damgaard:2019ztj, Huang:2021jlh, He:2021llb}).
In~\cite{He:2022cup}, it is argued that by projecting both external and loop momenta to $D=3$ of four-point amplituhedron in ${\cal N}=4$ sYM, we get the amplituhedron of four-point integrands in ${\cal N}=6$ Chern-Simons-matter theory, or ABJM theory~\cite{Hosomichi:2008jb, Aharony:2008ug}, and then the construction is generalized to any multiplicity in \cite{He:2023rou}. 

In \cite{Arkani-Hamed:2021iya}, the four-point amplituhedron of planar ${\cal N}=4$ sYM is decomposed into the so-called {\it negative geometries}, where the sum of all \text{connected} negative geometries naturally gives the integrand of logarithmic of the 4-point amplitude. Another important feature of this decomposition is that the infrared (IR) structure of integrated connected negative geometry is quite mild, it can only has at most $\epsilon^{-2}$ ($\epsilon^{-2L}$ for general $L$-loop diagrams) divergence after integrating all loop momenta in the dimensional regularization $D=4-2\epsilon$, which is in fact a one-loop divergence. Therefore, if we first fix one loop momentum not to integrate and integrate the other loop momenta in a connected negative geometry, we will get an IR finite object, which has the dual conformal symmetry after dividing a suitable factor, or equivalently which is a function in
\begin{equation}
z=\frac{(\ell-x_2)^2 (\ell-x_4)^2 (x_1-x_3)^2}{(\ell- x_1)^2 (\ell-x_3)^2 (x_2-x_4)^2}\,.
\end{equation}
where the dual momentum $x_i$ is defined as $x_{i+1}-x_i=p_i$ and $x_5=x_1$. For $\mathcal N=4$ sYM theory, this IR finite object is calculated up to $L=4$ \cite{Alday:2011ga,Engelund:2011fg,Engelund:2012re,Arkani-Hamed:2021iya,Alday:2012hy,Alday:2013ip,Henn:2019swt,Chicherin:2022bov}. It is also remarkable in \cite{Arkani-Hamed:2021iya} that a subset of all-loop negative geometries can be completely solved by differential equations relating these negative geometries.

For the four-point ABJM amplituhedron, we can also investigate the similar decomposition into negative geometries. In \cite{He:2022cup}, it shows that we can further decompose connected negative geometries according to the emergent causal structure, and then most of geometries cancel in the final decomposition, which leads a huge simplification comparing to the ${\cal N}=4$ sYM case. For example, the integrand of four-loop log of amplitude in ABJM theory is given by \textit{only} $4$ distinct negative geometries (with permutation) which is represent by \textit{bipartite} graphs
\begin{equation}\label{4labjm}
\tilde \Omega_4=-4\begin{tikzpicture}[baseline={([yshift=-.5ex]current bounding box.center)}]
\draw (-1.5,0) -- (-2,-1);
\draw (-1.5,0) -- (-2,-1);
\draw (-1.5,0) -- (-1.5,-1);
\draw (-1.5,0) -- (-1,-1);
\draw[fill=white] (-1,-1) circle[radius=.5ex];
\draw[fill=white] (-1.5,-1) circle[radius=.5ex];
\draw[fill=white] (-2,-1) circle[radius=.5ex];
\draw[fill=black] (-1.5,0) circle[radius=.5ex];
\end{tikzpicture}
-
4\begin{tikzpicture}[baseline={([yshift=-.5ex]current bounding box.center)}]
\draw (-1.5,0) -- (-2,-1);
\draw (-1.5,0) -- (-2,-1);
\draw (-1.5,0) -- (-1.5,-1);
\draw (-1.5,0) -- (-1,-1);
\draw[fill=black] (-1,-1) circle[radius=.5ex];
\draw[fill=black] (-1.5,-1) circle[radius=.5ex];
\draw[fill=black] (-2,-1) circle[radius=.5ex];
\draw[fill=white] (-1.5,0) circle[radius=.5ex];
\end{tikzpicture}
-
24\,\,\begin{tikzpicture}[baseline={([yshift=-.5ex]current bounding box.center)}]
\draw (-2,0) -- (-0.5,0);
\draw[fill=white] (-1,0) circle[radius=.5ex];
\draw[fill=black] (-1.5,0) circle[radius=.5ex];
\draw[fill=white] (-2,0) circle[radius=.5ex];
\draw[fill=black] (-0.5,0) circle[radius=.5ex];
\end{tikzpicture}
+
6\,\,\begin{tikzpicture}[baseline={([yshift=-.5ex]current bounding box.center)}]
\draw (-2,-1.5) -- (-1,-1.5) -- (-1,-0.5) -- (-2,-0.5) -- cycle;
\draw[fill=white] (-1,-1.5) circle[radius=.5ex];
\draw[fill=black] (-1,-0.5) circle[radius=.5ex];
\draw[fill=white] (-2,-0.5) circle[radius=.5ex];
\draw[fill=black] (-2,-1.5) circle[radius=.5ex];
\end{tikzpicture}\quad ,
\end{equation}
where each node of the diagram represents a loop.
For each ABJM negative geometry, it shares the nice IR structure as the ${\cal N}=4$ sYM negative geometry. 
Similarly, each negative geometry defines a IR finite object after integrating $L-1$ loop momenta out of $L$ loops, which is calculated up to $L=3$ \cite{He:2023exb,Henn:2023pkc}.


For both ${\cal N}=4$ sYM and ABJM theory, the above IR finite object is in fact related to the Wilson loop with a Lagrangian insertion. By Wilson loop-amplitude duality (see \textit{e.g.} \cite{Alday:2007hr, Alday:2007he, Alday:2009yn,Brandhuber:2007yx,Drummond:2007aua,Drummond:2007cf,Drummond:2007bm,Drummond:2008aq, Bern:2008ap,CaronHuot:2010ek, Mason:2010yk}), the four-point amplitudes is related to the polygon Wilson loop with 4 vertex $x_1,\dots,x_4$:
\begin{equation}\label{wlduality}
\log W_4(x_1,x_2,x_3,x_4) \sim \log (A_4/A_4^{(0)}),
\end{equation}
where $A_4^{(0)}$ is the tree-level amplitude, and this duality also holds at the integrand level. 
According to the Lagrangian insertion formula $g^2 \partial_{g^2}\langle\mathcal{O}\rangle=-i \int d^D \ell\left\langle\mathcal{O} \mathcal{L}\left(\ell\right)\right\rangle$ ($g^2$ is the expansion parameter), the important IR-finite quantity from integrating all but one loop of the negative geometry is related to the Wilson loop with a Lagrangian insertion~\cite{Engelund:2011fg, Alday:2013ip, Chicherin:2022bov, Chicherin:2022zxo}. 

The Wilson loop $\log W_4$ has ultraviolet (UV) divergence, which corresponds to propagating gluons between two adjacent sides of the Wilson loop around a cusp perturbatively. To regularize this divergence, one can introduce a renormalization factor $Z_{\text{cusp}}$, and therefore define the a cusp anomalous dimension as $\Gamma_{\text{cusp}}=\mu \partial_\mu Z_{\text{cusp}}$, where $\mu$ is the renormalization scale of the theory. The cusp anomalous dimension $\Gamma_{\text{cusp}}$ appears in $\log W_4$ as
\[
\log \left\langle W_4\right\rangle=-2 \sum_{L=1}^{\infty} \frac{\lambda^L \Gamma_{\text {cusp }}^{(L)}}{(L \epsilon)^2}+\mathcal{O}(\epsilon^{-1})
\]
in the dimensional regularization.  Therefore, both for the ${\cal N}=4$ sYM~\cite{Arkani-Hamed:2021iya} and the ABJM~\cite{He:2023exb,Henn:2023pkc} theory, we can extract the cusp anomalous dimension from the $\epsilon^{-2}$ order of IR divergence of the logarithm of the $4$-point amplitude by the Wilson loop-amplitude duality. Based this method, the two-loop cusp anomalous dimension of the ABJM theory is computed in \cite{He:2023exb,Henn:2023pkc}.
While as proposed in \cite{He:2023exb,Henn:2023pkc}, there is no contribution to the cusp anomalous dimension from negative geometry with an odd number of loops, so the next leading order of the cusp anomalous dimension is at $L=4$.


In this note, we are mainly interested in integrating three loop momenta out of $4$ loops for each negative geometry in eq.\eqref{4labjm} in ABJM theory, which can be used to calculate the $4$-loop cusp anomalous dimension $\Gamma_{\text{cusp}}^{(4)}$ of the ABJM theory. The note is organized as follows. In the next subsection, we will review ABJM four-point amplituhedron and negative geometries and gives the explicit integrand for each bipartite graph in eq.\eqref{4labjm}. In section \ref{sec:2}, we will integrate three loop momenta out of $4$ loops for each diagram based on Mellin method. In section \ref{sec:3}, we extract the cusp anomalous dimension by further integrating the last loop momentum, and we find that the box diagram 
\begin{equation}
\int\prod_{i=1}^4d^3\ell_i\begin{tikzpicture}[baseline={([yshift=-.5ex]current bounding box.center)},scale=0.75]
\draw (-2,-1.5) -- (-1,-1.5) -- (-1,-0.5) -- (-2,-0.5) -- cycle;
\draw[fill=white] (-1,-1.5) circle[radius=.5ex];
\draw[fill=black] (-1,-0.5) circle[radius=.5ex];
\draw[fill=white] (-2,-0.5) circle[radius=.5ex];
\draw[fill=black] (-2,-1.5) circle[radius=.5ex];
\node at (-2.2,-1.7) {$1$};
\node at (-0.8,-1.7) {$2$};
\node at (-0.8,-0.3) {$3$};
\node at (-2.2,-0.3) {$4$};
\end{tikzpicture}= O(\epsilon^{-1})
\end{equation}
does not contribute to the cusp anomalous dimension, and we prove it by looking at the integrand in collinear regions. In section \ref{sec:4}, we consider a subset of all-loop (ladder-type) negative geometries, and we present a recursion relation based Mellin method, and conjecture that this type of negative geometries only contributes the power of $\pi$ and no other zeta values like $\zeta_3,\zeta_5$ to the cusp anomalous dimension.

\subsection{Review of ABJM four-point amplituhedron and negative geometries} 

Roughly speaking, the ABJM amplituhedron is defined by reducing the dimension of the 4 dimensional $\mathcal N=4$ sYM amplituhedron. The reducing is archived by imposing a universal projection for each external and loop momentum. For imposing this kinematics condition, we first review the kinematics of both $\mathcal N=4$ sYM and ABJM theories and fix the corresponding notations.

For planar massless theories with external momentum $\{p_i\,:\,1\leq i\leq n\}$, it is convenient to introduce dual momentum (or dual point) $\{x_i\,:\,1\leq i\leq n\}$, defined through $p_i={x}_{i+1}-{x}_i$ (with $x_{n+1}=x_1$)~\cite{Chen:2011vv}, and any physical quantities in these theories are functions in planar kinematics variables
\begin{equation}
(x_i-x_j)^2=(p_i+p_{i+1}+\cdots+p_{j-1})^2.
\end{equation}

To linearize the conformal group of dual space (dual conformal group), we embed the dual space $\mathbb R^{d,1}$ into $\mathbb R^{d+1,2}$ with two
extra dimensions \cite{Caron-Huot:2014lda,Simmons-Duffin:2012juh}, where the dual conformal group becomes $\operatorname{SO}(d+1,2)$. Each dual momentum $x^\mu$ is sent to a vector $X=(x^2,1,x^\mu)\in \mathbb R^{d+1,2}$,
and the metric of $\mathbb R^{d+1,2}$ is
\[
\eta=\begin{pmatrix}
0&1&\\
1&0&\\
&&-2\eta^{\mu\nu}
\end{pmatrix},
\]
where $\eta^{\mu\nu}$ is the spacetime metric. The basic $\operatorname{SO}(4,2)$ invariants are inner products of two vectors in the embedding space. We denote the inner product of two vectors $X$ and $Y$ as $(X\cdot Y)$, and define $X_i:=(x_i^2,1,x_i^\mu)$ and the shorthand 
\begin{equation}
(i\cdot j):=(X_i\cdot X_j)=(x_i-x_j)^2.
\end{equation}
Since the square of $X=(x^2,1,x^\mu)$ vanishes, \textit{i.e.} $(X\cdot X)=x^2-x^2=0$, the dual space can be identified more intrinsically as a quadric defined by
$(X\cdot X)=0$ module the identification $X \sim \alpha X$ for $\alpha\neq0$. Dual conformal invariants (DCI) of dual momentum should be built from inner product of these $X$'s, and it can only be the function of cross-ratios of inner products.

For the four-dimensional spacetime, such $X$ can be further parameterized by a wedge product of two twistors (a bi-twistor) $A\wedge B:=(AB)$ in a twistor space $\mathbb P^3$, such twistors call called momentum twistors~\cite{Hodges:2009hk}, and the dimension of space of bi-twistors is also ${4\choose 2}-2=6-2=4$. 
The inner product now corresponds to the wedge product,
\[
(X\cdot X')\propto (AB)\wedge (A'B')=A\wedge B\wedge A'\wedge B'=:\langle ABA'B'\rangle,
\]
and the proportionality factors will finally cancel in any dual conformal invariants. Since $p_i=x_{i+1}-x_i$ is massless, or $(i\cdot i+1)=0$, each $X_i$ can be choosen as a bi-twistor $(Z_{i-1}Z_i)$ (with $Z_0=Z_{n}$) for $n$ momentum twistors $\{Z_i\}_{i=1,\dots,n}$, and each loop momentum $\ell_i$ corresponds to a bi-twistor $(A_iB_i)$. We also introduce a shorthand 
\begin{equation}
\langle ijkl\rangle:=\langle Z_iZ_jZ_kZ_l\rangle,\quad 
\langle \ell_ikl\rangle:=\langle A_iB_iZ_kZ_l\rangle,
\quad \langle \ell_i\ell_j\rangle:=\langle A_iB_iA_jB_j\rangle.
\end{equation}

However, for the three-dimensional spacetime, we cannot view a dual momentum $X$ as a bi-twistor in any twistor space $\mathbb P^k$, the dimension ${k\choose 2}-2$ will never be $3$. In~\cite{Elvang:2014fja}, a nice way to represent 3D kinematics by taking a projection from 4D momentum twistor was proposed as follows. We first go back the the embedding space, by setting a vector $\Psi$ with $(\Psi\cdot \Psi)=1$, we require that all dual points and loop momentum are orthogonal to $\Psi$, \textit{i.e.}
\[
(X_i\cdot \Psi)=0,\quad (\ell\cdot \Psi)=0,
\]
and then the conformal group is naturally reduced to $\operatorname{SO}(2,3)$. In twistor space, this is equivalent to impose a {\it symplectic condition} on both external bi-twistor $(Z_a Z_{a{+}1})$ and loop variables $\ell_i\sim (A_iB_i)$ that 
\begin{equation}\label{sympletic}
\Psi_{IJ} Z_a^I Z_{a{+}1}^J=\Psi_{IJ} A_i^I B_i^J=0\,, \quad \text{with a choosen}\quad \Psi=
\begin{pmatrix}
&&&1\\
&&-1&\\
&1&&\\
-1&&&
\end{pmatrix}
\end{equation}
where $I=1,\dots,4$ is the index of the twistor space $\mathbb P^3$.

Now we can give the definition of the ABJM amplituhedron for four points \cite{He:2022cup}, the generalization to $n$ points see \cite{He:2023rou}. The four-point ABJM amplituhedron is defined by 
\begin{equation}\label{ineq}
\langle \ell_i 12\rangle, \langle \ell_i 23\rangle, \langle \ell_i 34\rangle, \langle \ell_i 14\rangle<0, \langle \ell_i 13\rangle, \langle \ell_i 24\rangle>0, \langle \ell_i \ell_j\rangle<0,  
\end{equation}
all defined on the support of eq.\eqref{sympletic}, 
and its canonical form will produce the all-loop ABJM integrand. Note that, there is an important subtlety is that $\langle 1234\rangle<0$ for real $Z$'s satisfying the symplectic condition,
so we need to flip the overall sign for the definition of the $D=4$ amplituhedron~\cite{Arkani-Hamed:2013jha}. 

When computing the canonical form, $\langle \ell_i \ell_j\rangle<0$ is the most non-trivial condition which mix different loops. In~\cite{Arkani-Hamed:2021iya}, there is a new decomposition for the four-point amplituhedron~\cite{Arkani-Hamed:2013kca} based on a simple fact that 
\[
\Omega(C,\langle \ell_i \ell_j\rangle>0)+\Omega(C,\langle \ell_i \ell_j\rangle<0)=\Omega(C),
\]
where $\Omega$ is the canonical form and $C$ is the set of the other inequalities. 
Therefore, just replacing $\Omega(C_j,\langle \ell_i \ell_j\rangle<0)$ by $\Omega(C)-\Omega(C_i,\langle \ell_i \ell_j\rangle>0)$ for all possible $\langle \ell_i\ell_j\rangle$, we can decompose the geometry to geometries with only ``mutual negativity" conditions $\langle \ell_i \ell_j\rangle>0$ such that the $L$-loop geometry $\mathcal A_L$ can be decomposed into  
\begin{equation}
{\cal A}_L=\sum_g (-)^{E(g)} {\cal A}(g),
\end{equation} 
where we use a graph $g$ with $L$ nodes to label a geometry $\mathcal A(g)$ and sum over all possible (maybe disconnected) graphs without $2$-cycles. The node of a graph denote a loop, and a edge between two nodes $i$ and $j$ means we need to apply the condition $\langle \ell_i \ell_j\rangle>0$. For each graph, we need less conditions $\langle \ell_i \ell_j\rangle>0$, so it will be easier to compute its canonical form. Furthermore, it suffices to consider all {\it connected} graphs, whose (signed) sum gives the geometry for the logarithm of amplitudes~\cite{Arkani-Hamed:2021iya}.

A further simplification \cite{He:2022cup} occurs in 3D ABJM amplitudedron, most of these geometries do not contribute at all. We can further decompose $\langle \ell_i \ell_j\rangle>0$ into two regions  which gives two possible directions on each edge, and we need to sum over all possible directed graphs.
In such decomposition, the condition given by $i\to j\to k$ implies the condition $i\to k$, so different graphs can represent the same geometry, which leads a huge cancellation in the signed summation. Finally, we only need to sum over all \textit{bipartite} graphs. A bipartite graph is a directed graph with only sinks and sources, we will use black/white node to denote a source/sink vertex and omit all directions on edges in a bipartite. Therefore, the canonical form of $L$-loop amplitude $\Omega_L$ and log of $L$-loop amplitude $\tilde\Omega_L$ read
\begin{equation}
\Omega_L=\sum_{\text{bipartite }g} (-)^{E(g)} \Omega(g),\quad\tilde\Omega_L=\sum_{\substack{\text{connected}\\\text{bipartite }g}} (-)^{E(g)} \Omega(g).
\end{equation}

For example, for the 3-loop case, only the chain graph contributes, {\it i.e.}
\begin{equation}\label{eq:3loopg}
    \tilde\Omega_3=
\begin{tikzpicture}[baseline={([yshift=-1.7ex]current bounding box.center)}]
\draw (-2,0) -- (-1,0);
\draw[fill=black] (-1,0) circle[radius=.5ex];
\draw[fill=white] (-1.5,0) circle[radius=.5ex];
\draw[fill=black] (-2,0) circle[radius=.5ex];
\node at (-2,0.3) {$1$};
\node at (-1.5,0.3) {$2$};
\node at (-1,0.3) {$3$};
\end{tikzpicture}+5\,\text{permutations},
\end{equation}
and only the two kinds of tree graphs and the box contribute for $\tilde\Omega_4$, {\it i.e.}
\begin{equation}
\tilde \Omega_4=-4\begin{tikzpicture}[baseline={([yshift=-.5ex]current bounding box.center)}]
\draw (-1.5,0) -- (-2,-1);
\draw (-1.5,0) -- (-2,-1);
\draw (-1.5,0) -- (-1.5,-1);
\draw (-1.5,0) -- (-1,-1);
\draw[fill=white] (-1,-1) circle[radius=.5ex];
\draw[fill=white] (-1.5,-1) circle[radius=.5ex];
\draw[fill=white] (-2,-1) circle[radius=.5ex];
\draw[fill=black] (-1.5,0) circle[radius=.5ex];
\end{tikzpicture}
-
4\begin{tikzpicture}[baseline={([yshift=-.5ex]current bounding box.center)}]
\draw (-1.5,0) -- (-2,-1);
\draw (-1.5,0) -- (-2,-1);
\draw (-1.5,0) -- (-1.5,-1);
\draw (-1.5,0) -- (-1,-1);
\draw[fill=black] (-1,-1) circle[radius=.5ex];
\draw[fill=black] (-1.5,-1) circle[radius=.5ex];
\draw[fill=black] (-2,-1) circle[radius=.5ex];
\draw[fill=white] (-1.5,0) circle[radius=.5ex];
\end{tikzpicture}
-
24\,\,\begin{tikzpicture}[baseline={([yshift=-.5ex]current bounding box.center)}]
\draw (-2,0) -- (-0.5,0);
\draw[fill=white] (-1,0) circle[radius=.5ex];
\draw[fill=black] (-1.5,0) circle[radius=.5ex];
\draw[fill=white] (-2,0) circle[radius=.5ex];
\draw[fill=black] (-0.5,0) circle[radius=.5ex];
\end{tikzpicture}
+
6\,\,\begin{tikzpicture}[baseline={([yshift=-.5ex]current bounding box.center)}]
\draw (-2,-1.5) -- (-1,-1.5) -- (-1,-0.5) -- (-2,-0.5) -- cycle;
\draw[fill=white] (-1,-1.5) circle[radius=.5ex];
\draw[fill=black] (-1,-0.5) circle[radius=.5ex];
\draw[fill=white] (-2,-0.5) circle[radius=.5ex];
\draw[fill=black] (-2,-1.5) circle[radius=.5ex];
\end{tikzpicture}
\end{equation}
This simplification becomes more significant with $L$ increasing. For $L=2,\dots, 7$, the number of topologies for connected graphs are $1,2,6,21,112, 853$, but that of bipartite topologies decrease to $1,1,3, 5, 17, 44$. All canonical forms of negative geometries up to $L=5$ have been computed~\cite{He:2022cup}. We here list all graphs and their integrands up to $L=4$,
\begin{align}
&\begin{tikzpicture}[baseline={([yshift=-1.7ex]current bounding box.center)}]
\draw[fill=black] (-2,0) circle[radius=.5ex];
\node at (-2,0.3) {$1$};
\end{tikzpicture}=\frac{c\epsilon_1}{s_1 t_1},\quad \begin{tikzpicture}[baseline={([yshift=-1.7ex]current bounding box.center)}]
\draw (-2,0) -- (-1.5,0);
\draw[fill=black] (-2,0) circle[radius=.5ex];
\draw[fill=white] (-1.5,0) circle[radius=.5ex];
\node at (-2,0.3) {$1$};
\node at (-1.5,0.3) {$2$};
\end{tikzpicture}=\frac{2 c^2}{s_1 t_2D_{1,2}},
\quad \begin{tikzpicture}[baseline={([yshift=-1.7ex]current bounding box.center)}]
\draw (-2,0) -- (-1,0);
\draw[fill=black] (-1,0) circle[radius=.5ex];
\draw[fill=white] (-1.5,0) circle[radius=.5ex];
\draw[fill=black] (-2,0) circle[radius=.5ex];
\node at (-2,0.3) {$1$};
\node at (-1.5,0.3) {$2$};
\node at (-1,0.3) {$3$};
\end{tikzpicture}=\frac{4 c^2 \epsilon_2 }{s_1 t_2 s_3 D_{1,2} D_{2,3}}\\
&\begin{tikzpicture}[baseline={([yshift=-1.7ex]current bounding box.center)}]
\draw (-2,0) -- (-0.5,0);
\draw[fill=black] (-1,0) circle[radius=.5ex];
\draw[fill=white] (-1.5,0) circle[radius=.5ex];
\draw[fill=black] (-2,0) circle[radius=.5ex];
\draw[fill=white] (-0.5,0) circle[radius=.5ex];
\node at (-2,0.3) {$1$};
\node at (-1.5,0.3) {$2$};
\node at (-1,0.3) {$3$};
\node at (-0.5,0.3) {$4$};
\end{tikzpicture}=
\frac{8 c^2\epsilon_2 \epsilon_3}{D_{1,2} D_{2,3} D_{3,4} s_1  t_2 s_3 t_4},\quad 
\begin{tikzpicture}[baseline={([yshift=-.5ex]current bounding box.center)},scale=0.75]
\draw (-1.5,0) -- (-2,-1);
\draw (-1.5,0) -- (-2,-1);
\draw (-1.5,0) -- (-1.5,-1);
\draw (-1.5,0) -- (-1,-1);
\draw[fill=white] (-1,-1) circle[radius=.5ex];
\draw[fill=white] (-1.5,-1) circle[radius=.5ex];
\draw[fill=white] (-2,-1) circle[radius=.5ex];
\draw[fill=black] (-1.5,0) circle[radius=.5ex];
\node at (-1.5,0.3) {$1$};
\node at (-2.2,-1.3) {$2$};
\node at (-1.5,-1.3) {$3$};
\node at (-0.8,-1.3) {$4$};
\end{tikzpicture}
=\frac{8 c^3  t_1}{D_{1,2} D_{1,3} D_{1,4} s_1 t_2 t_3 t_4}
\label{l4neg1}\\
&\begin{tikzpicture}[baseline={([yshift=-.5ex]current bounding box.center)},scale=0.75]
\draw (-2,-1.5) -- (-1,-1.5) -- (-1,-0.5) -- (-2,-0.5) -- cycle;
\draw[fill=white] (-1,-1.5) circle[radius=.5ex];
\draw[fill=black] (-1,-0.5) circle[radius=.5ex];
\draw[fill=white] (-2,-0.5) circle[radius=.5ex];
\draw[fill=black] (-2,-1.5) circle[radius=.5ex];
\node at (-2.2,-1.7) {$1$};
\node at (-0.8,-1.7) {$2$};
\node at (-0.8,-0.3) {$3$};
\node at (-2.2,-0.3) {$4$};
\end{tikzpicture}=4\frac{4 \epsilon_1 \epsilon_2 \epsilon_3 \epsilon_4- 
c (\epsilon_1 \epsilon_3 N^t_{24} + \epsilon_2 \epsilon_4 N^s_{13})- 
c^2 N^{\rm cyc}_{1,2,3,4} }{D_{1,2} D_{2,3} D_{3,4} D_{4,1} s_1 t_2 s_3 t_4}
\label{l4neg2}
\end{align}
where 
\[
D_{i,j}=-\langle \ell_i\ell_j\rangle,\,s_i=\langle \ell_i 12\rangle \langle \ell_i 34\rangle,\, t_i=\langle \ell_i 23\rangle \langle \ell_i 14\rangle,\, c=\langle 1234\rangle,\, 
\epsilon_i=\sqrt{\langle 1234\rangle \langle \ell_i 13\rangle \langle \ell_i 24\rangle}
\]
and 
\[
\begin{aligned}
  &N^s_{13}:=\langle \ell_1 12\rangle \langle \ell_3 34\rangle + \langle \ell_3 12\rangle \langle \ell_1 34\rangle\\\nonumber
  &N^t_{24}:=-\langle \ell_2 41\rangle \langle \ell_4 23\rangle - \langle \ell_4 41\rangle \langle \ell_2 23\rangle\\\nonumber
  &N^{\rm cyc}_{i,j,k,l}:=\langle \ell_i 1 2\rangle  \langle \ell_j 3 4\rangle\langle \ell_k 1 2\rangle\langle \ell_l 3 4\rangle + {\rm cyc}(1,2,3,4).
  \end{aligned}
\]
Other graphs with white and black vertex interchanged can be obtained by $s\leftrightarrow t$ (or equivalently $X_1\leftrightarrow X_2,X_3 \leftrightarrow X_4$).

To translate expressions in momentum twistors to embedding space, we give a dictionary here:
\begin{equation}
\begin{aligned}
&\frac{\epsilon_i}c\to \frac{\epsilon(1,2,3,4,\ell_i)}{-(1\cdot 3)(2\cdot 4)},\quad 
\frac{s_i}{c}\to \frac{(\ell_i\cdot 2)(\ell_i\cdot 4)}{(2\cdot 4)},\quad \frac{t_i}{c}\to \frac{(\ell_i\cdot 1)(\ell_i\cdot 3)}{(1\cdot 3)},\\
&D_{i,j}\to -(\ell_i\cdot \ell_j),\quad 
c\to -(1\cdot 3) \text{ or } (2\cdot 4),
\end{aligned}
\end{equation}
where $\epsilon$ is the totally anti-symmetric tensor in the embedding space $\mathbb R^{2,3}$:
\[
\epsilon(1,2,3,4,\ell_i):=\epsilon_{IJKLM}X_1^IX_2^JX_3^KX_4^L(\ell_i)^M,
\]
which is normalized as 
\[\epsilon(y_1,y_2,y_3,y_4,y_5)\epsilon(z_1,z_2,z_3,z_4,z_5)=-\det((y_i\cdot z_j))/2,
\]
and we need to choose $c$ to be $-(1\cdot 3)$ or $(2\cdot 4)$ to make integrals dual conformal invariant in embedding space.

In this note, we focus on calculating the finite quantity of ABJM four-point amplituhedron
\begin{equation}
{\cal W}_L(\ell, 1,2,3,4):=\int \prod_{i=2}^L \frac{d^3 \ell_i}{i\pi^{3/2}} \, \tilde{\Omega}_L,  
\end{equation}
which depends on the last loop $\ell_1$ and external points; after stripping off a non-DCI prefactor it becomes function of a single cross-ratio
\begin{equation}\label{defz}
z=\frac{(\ell\cdot 2) (\ell\cdot 4) (1\cdot 3)}{(\ell\cdot 1) (\ell\cdot 3) (2\cdot 4)}\,.
\end{equation}
This will be the main target of loop integrations for ABJM amplituhedron.
 
\section{Integrating the $4$-loop ABJM negative geometries}\label{sec:2}

In this section, we consider the corresponding ${\cal W}_L(\ell, 1,2,3,4)$ for $L=4$ negative geometries whose integrands are listed in eq.\eqref{l4neg1} and \eqref{l4neg2}, and the log of amplitude is given by
\begin{equation}
\tilde \Omega_4=-4\begin{tikzpicture}[baseline={([yshift=-.5ex]current bounding box.center)}]
\draw (-1.5,0) -- (-2,-1);
\draw (-1.5,0) -- (-2,-1);
\draw (-1.5,0) -- (-1.5,-1);
\draw (-1.5,0) -- (-1,-1);
\draw[fill=white] (-1,-1) circle[radius=.5ex];
\draw[fill=white] (-1.5,-1) circle[radius=.5ex];
\draw[fill=white] (-2,-1) circle[radius=.5ex];
\draw[fill=black] (-1.5,0) circle[radius=.5ex];
\end{tikzpicture}
-
4\begin{tikzpicture}[baseline={([yshift=-.5ex]current bounding box.center)}]
\draw (-1.5,0) -- (-2,-1);
\draw (-1.5,0) -- (-2,-1);
\draw (-1.5,0) -- (-1.5,-1);
\draw (-1.5,0) -- (-1,-1);
\draw[fill=black] (-1,-1) circle[radius=.5ex];
\draw[fill=black] (-1.5,-1) circle[radius=.5ex];
\draw[fill=black] (-2,-1) circle[radius=.5ex];
\draw[fill=white] (-1.5,0) circle[radius=.5ex];
\end{tikzpicture}
-
24\,\,\begin{tikzpicture}[baseline={([yshift=-.5ex]current bounding box.center)}]
\draw (-2,0) -- (-0.5,0);
\draw[fill=white] (-1,0) circle[radius=.5ex];
\draw[fill=black] (-1.5,0) circle[radius=.5ex];
\draw[fill=white] (-2,0) circle[radius=.5ex];
\draw[fill=black] (-0.5,0) circle[radius=.5ex];
\end{tikzpicture}
+
6\,\,\begin{tikzpicture}[baseline={([yshift=-.5ex]current bounding box.center)}]
\draw (-2,-1.5) -- (-1,-1.5) -- (-1,-0.5) -- (-2,-0.5) -- cycle;
\draw[fill=white] (-1,-1.5) circle[radius=.5ex];
\draw[fill=black] (-1,-0.5) circle[radius=.5ex];
\draw[fill=white] (-2,-0.5) circle[radius=.5ex];
\draw[fill=black] (-2,-1.5) circle[radius=.5ex];
\end{tikzpicture}\quad .
\end{equation}

To perform the integration of loop momentum and consider the resummation, we follow the normalization in \cite{Henn:2023pkc}. The expansion of the $\log$ of the integrand with respect to {}'t Hooft coupling $\lambda=N/k$ is 
\begin{equation}
\mathcal L=\sum_{L=1}^\infty\lambda^L \mathcal L_L=\sum_{L=1}^\infty\frac{1}{L!}\biggl(\frac{i}{2\sqrt{\pi}}\biggr)^{L} \lambda^{L}\tilde\Omega_L=\sum_{g}\frac{1}{L_g!}\biggl(\frac{i}{2\sqrt{\pi}}\biggr)^{L_g} \lambda^{L_g}(-1)^{E_g}\tilde\Omega_g,
\end{equation}
where we sum over all connected bipartite graphs $g$ with $L_g$ nodes and $E_g$ edges in the last summation. After performing the loop integration of $\ell_2,\dots$, we define 
\begin{equation}
F_g=\frac{1}{L_g!}\biggl(\frac{i}{2\sqrt{\pi}}\biggr)^{L_g} \biggl[\frac{1}{\sqrt{\pi}}\biggl(\frac{(1\cdot 3)(2\cdot 4)}{(\ell\cdot 1)(\ell\cdot 3)(\ell\cdot 2)(\ell\cdot 4)}\biggr)^{3/4}\biggr]^{-1}\int \prod_{i=2}^L \frac{d^3 \ell_i}{i\pi^{3/2}} \, \tilde{\Omega}_g,
\end{equation}
where the prefactor is introduced to cancel the dual conformal weight of the left loop momentum such that $F_g$ is a dual conformal invariant\footnote{The prefactor for the negative geometry with an odd number of loops is different, see \cite{He:2023exb,Henn:2023pkc}.}, or is a function in $z$.

We will separate the calculation into three type of diagrams in three subsections: star, box and ladder. We here first collect all results:
\begin{align}
F_{\text{star}}&=-\frac{\pi^{3}}{12}(z^{-1/4}+z^{1/4}),\\
F_{\text{box}}&=\frac{\pi^3}{32}\biggl(\frac{2}{3}z^{-1/4}+2z^{1/4}-\frac{4}{\pi^2}z^{-1/4}(1+z)f(z)\biggr)+(z\leftrightarrow 1/z),\\
F_{\text{ladder}}&=-\frac{\pi^3}{8}z^{-1/4}\biggl(1+\frac{\log(z/4)^2}{2\pi^2}+\frac{(1+z)f(z)}{\pi^2}\biggr)+(z\leftrightarrow z^{-1}),
\end{align}
where $f(z)$ is a weight-$2$ polylogarithm defined in eq.\eqref{wfz}.

\subsection{Star diagram}

The integrand for star diagram
\begin{equation}
\begin{aligned}
S=\begin{tikzpicture}[baseline={([yshift=-.5ex]current bounding box.center)}]
\draw (-1.5,0) -- (-2,-1);
\draw (-1.5,0) -- (-2,-1);
\draw (-1.5,0) -- (-1.5,-1);
\draw (-1.5,0) -- (-1,-1);
\draw[fill=white] (-1,-1) circle[radius=.5ex];
\draw[fill=white] (-1.5,-1) circle[radius=.5ex];
\draw[fill=white] (-2,-1) circle[radius=.5ex];
\draw[fill=black] (-1.5,0) circle[radius=.5ex];
\node at (-1.5,0.3) {$1$};
\node at (-2.2,-1.3) {$2$};
\node at (-1.5,-1.3) {$3$};
\node at (-0.8,-1.3) {$4$};
\end{tikzpicture}
&=8 c^3 \frac{t_1}{D_{1,2} D_{1,3} D_{1,4} s_1 t_2 t_3 t_4}\\
&=-8\frac{(\ell_1\cdot 1)(\ell_1\cdot 3)\inp24}{(\ell_1\cdot 2)(\ell_1\cdot 4)\inp13}\prod_{i=2}^4\frac{\inp13}{(\ell_1\cdot \ell_i)(\ell_i\cdot 1)(\ell_i\cdot 3)},
\end{aligned}
\end{equation}
we further use a subscript to denote which node is left to unintegrated, 
\[
S_i=\biggl[\frac{1}{\sqrt{\pi}}\biggl(\frac{(2\cdot 4)(1\cdot 3)}{(\ell\cdot 2)(\ell_1\cdot 4)(\ell\cdot 1)(\ell\cdot 3)}\biggr)^{3/4}\biggr]^{-1}\int_{\{1,2,3,4
\}\setminus \{i\}}S,
\]
and unlabeled $\ell$ or $z$ always corresponds to unintegrated node.

It is easy to perform the loop integral for $S$ by recursively using the one-loop scale triangle integral in 3D
\begin{equation}\label{ost}
\int_{\ell} \frac{1}{(\ell\cdot a)(\ell\cdot b)(\ell \cdot c)}=\frac{\pi^{3/2}}{\sqrt{(a\cdot b)(b\cdot c)(c\cdot a)}},
\end{equation}
and we get that
\[
S_1=-8\pi^{5}z^{-1/4},\quad S_2=S_3=S_4=-8\pi^{5}z^{1/4}.
\]
and  
\[
F_S=4\begin{tikzpicture}[baseline={([yshift=-.5ex]current bounding box.center)}]
\draw (-1.5,0) -- (-2,-1);
\draw (-1.5,0) -- (-2,-1);
\draw (-1.5,0) -- (-1.5,-1);
\draw (-1.5,0) -- (-1,-1);
\draw[fill=white] (-1,-1) circle[radius=.5ex];
\draw[fill=white] (-1.5,-1) circle[radius=.5ex];
\draw[fill=white] (-2,-1) circle[radius=.5ex];
\draw[fill=black] (-1.5,0) circle[radius=.5ex];
\end{tikzpicture}=\frac{1}{4!}\biggl(\frac{i}{2\sqrt{\pi}}\biggr)^{4}(S_1+S_2+S_3+S_4)=-\frac{1}{48}\pi^{5}(z^{-1/4}+3z^{1/4}).
\]

Similarly, the contribution from the other star diagram
\begin{equation}
\bar S=\begin{tikzpicture}[baseline={([yshift=-.5ex]current bounding box.center)}]
\draw (-1.5,0) -- (-2,-1);
\draw (-1.5,0) -- (-2,-1);
\draw (-1.5,0) -- (-1.5,-1);
\draw (-1.5,0) -- (-1,-1);
\draw[fill=black] (-1,-1) circle[radius=.5ex];
\draw[fill=black] (-1.5,-1) circle[radius=.5ex];
\draw[fill=black] (-2,-1) circle[radius=.5ex];
\draw[fill=white] (-1.5,0) circle[radius=.5ex];
\node at (-1.5,0.3) {$1$};
\node at (-2.2,-1.3) {$2$};
\node at (-1.5,-1.3) {$3$};
\node at (-0.8,-1.3) {$4$};
\end{tikzpicture}
\end{equation}
is simply given by taking $z\to 1/z$ for $S$, so
\[
F_{\bar S}=-\frac{1}{48}\pi^{5}(z^{1/4}+3z^{-1/4}).
\]
The full star contribution is just
\begin{equation}
F_{\text{star}}=F_S+F_{\bar S}=-\frac{\pi^{3}}{12}(z^{-1/4}+z^{1/4}).
\end{equation}

\subsection{Box diagram}

The integration of 4-loop box diagram is much more difficult than the star diagram. The integrand is 
$$
B=\begin{tikzpicture}[baseline={([yshift=-.5ex]current bounding box.center)},scale=0.75]
\draw (-2,-1.5) -- (-1,-1.5) -- (-1,-0.5) -- (-2,-0.5) -- cycle;
\draw[fill=white] (-1,-1.5) circle[radius=.5ex];
\draw[fill=black] (-1,-0.5) circle[radius=.5ex];
\draw[fill=white] (-2,-0.5) circle[radius=.5ex];
\draw[fill=black] (-2,-1.5) circle[radius=.5ex];
\node at (-2.2,-1.7) {$1$};
\node at (-0.8,-1.7) {$2$};
\node at (-0.8,-0.3) {$3$};
\node at (-2.2,-0.3) {$4$};
\end{tikzpicture}=4\frac{4 \epsilon_1 \epsilon_2 \epsilon_3 \epsilon_4- 
c (\epsilon_1 \epsilon_3 N^t_{24} + \epsilon_2 \epsilon_4 N^s_{13})- 
c^2 N^{\rm cyc}_{1,2,3,4} }{D_{1,2} D_{2,3} D_{3,4} D_{4,1} s_1 t_2 s_3 t_4}
$$
where
\[
\begin{aligned}
  &N^s_{13}:=\langle \ell_1 12\rangle \langle \ell_3 34\rangle + \langle \ell_3 12\rangle \langle \ell_1 34\rangle\\\nonumber
  &N^t_{24}:=-\langle \ell_2 41\rangle \langle \ell_4 23\rangle - \langle \ell_4 41\rangle \langle \ell_2 23\rangle\\\nonumber
  &N^{\rm cyc}_{i,j,k,l}:=\langle \ell_i 1 2\rangle  \langle \ell_j 3 4\rangle\langle \ell_k 1 2\rangle\langle \ell_l 3 4\rangle + {\rm cyc}(1,2,3,4),
  \end{aligned}
\]
where $\text{cyc}(1,2,3,4)$ indicates cyclic rotations of dual points $12\to 23 \to 34 \to 41$. We want to calculate 
$$
B_1=\biggl(\frac{(2\cdot 4)\sqrt{(1\cdot 3)}}{(\ell_1\cdot 2)(\ell_1\cdot 4)\sqrt{(\ell_1\cdot 1)(\ell_1\cdot 3)}}\biggr)^{-1}\int_{\ell_2,\ell_3,\ell_4}B,
$$
where the new prefactor is just
\begin{equation}\label{eq:pref}
\begin{aligned}
\frac{(2\cdot 4)\sqrt{(1\cdot 3)}}{(\ell\cdot 2)(\ell\cdot 4)\sqrt{(\ell\cdot 1)(\ell\cdot 3)}}&=\biggl(\frac{(1\cdot 3)(2\cdot 4)}{(\ell\cdot 1)(\ell\cdot 3)(\ell\cdot 2)(\ell\cdot 4)}\biggr)^{3/4}z^{-1/4}.
\end{aligned}
\end{equation}

We first expand the integrand and integrate loop 2 and loop 4, which is symmetric and gives the same result. There're two types of 1-loop integrals when integrating loop 2 and 4, the one is scale triangle 
$$
\int_{\ell} \frac{1}{(\ell\cdot a)(\ell\cdot b)(\ell \cdot c)}=\frac{\pi^{3/2}}{\sqrt{(a\cdot b)(b\cdot c)(c\cdot a)}}
$$
and the other is a one-loop box integral with a numerator $N[\ell]$, 
$$
\int_{\ell} \frac{N[\ell]}{(\ell\cdot \ell_1)(\ell\cdot \ell_3)(\ell \cdot 1)(\ell\cdot 3)}=\pi^{3/2}\frac{N[\ell_1]/\sqrt{(\ell_1\cdot 1)(\ell_1\cdot 3)}+N[\ell_3]/\sqrt{(\ell_3\cdot 1)(\ell_3\cdot 3)}}{\sqrt{(1\cdot 3)(\ell_1\cdot \ell_3)}+\sqrt{(\ell_1\cdot 3)(\ell_3\cdot 1)}+\sqrt{(\ell_1\cdot 1)(\ell_3\cdot 3)}},
$$
where $N[\ell]$ is linear in dual momentum $\ell$ and $N[1]=N[3]=0$, for example, $N[\ell]$ could be $\epsilon_\ell$ or $(\ell\cdot 2)$ or $(\ell\cdot 4)$. Therefore, with expanding the numerator by
\[\epsilon(y_1,y_2,y_3,y_4,y_5)\epsilon(z_1,z_2,z_3,z_4,z_5)=-\det((y_i\cdot z_j))/2,
\]
we get that 
$$
\int_{\ell_2,\ell_4}B=\frac{ P}{H^2}+Q,
$$
where $P,Q$ are rational functions of scale products and $\epsilon$'s (\textit{i.e.} they can expand as sums of monomials of scale products and $\epsilon$'s), and 
\[H:=\sqrt{(1\cdot 3)(\ell_1\cdot \ell_3)}+\sqrt{(\ell_1\cdot 3)(\ell_3\cdot 1)}+\sqrt{(\ell_1\cdot 1)(\ell_3\cdot 3)}.\]
A monomial of scale products and $\epsilon$'s is just a 1-loop integrand and is easy to work out, so we can calculate $\int_{\ell_3}Q$. 

For calculating $\int_{\ell_3}P/H^2$, we introduce a inverse Mellin transform (which can be seen as an inverse Feynman parameterization) for $1/H^2$ 
\[
\frac{1}{(A+B+C)^2}=\int_{\gamma_u,\gamma_v} \frac{du}{2\pi i}\frac{dv}{2\pi i}A^{-u} B^{-v}C^{-w} \frac{\Gamma (u) \Gamma (v) \Gamma (w)}{\Gamma(u+v+w)},
\]
where $u+v+w=2$, and contours $\gamma_u,\gamma_v$ are defined that $\operatorname{Re}(u)>0,\operatorname{Re}(v)>0,\operatorname{Re}(w)>0$. Now we can first perform the $\ell_3$-integral for $P/H^2$ and then consider the inverse Mellin transform, the $\ell_3$-integrand is a sum of 1-loop integrals in the following form
$$
\int_{\ell} \frac{N[\ell]^{n_0}}{(\ell\cdot 1)^{n_1}(\ell\cdot 2)^{n_2}(\ell \cdot 3)^{n_3}(\ell \cdot 4)^{n_4}(\ell\cdot \ell_1)^{n_5}},
$$
where $n_0=0,1$ and $n_1+\cdots+n_5=3+n_0$, so we can introduce Feynman parameters $a_1,\dots,a_5$ ($a_5$ for $(\ell\cdot \ell_1)$) for each propagator and perform the loop integral. The result of this integral can be found in the Appendix.

Therefore, we get that 
\begin{equation}
\begin{aligned}
\frac{\pi^{-9/2}}{4}&\biggl(\frac{(2\cdot 4)\sqrt{(1\cdot 3)}}{(\ell_1\cdot 2)(\ell_1\cdot 4)\sqrt{(\ell_1\cdot 1)(\ell_1\cdot 3)}}\biggr)^{-1}\int_{\ell_2,\ell_3,\ell_4}B\\
&=\,-\frac{2}{\sqrt{z}}+\frac{2}{3}+(1+z)\biggl(\frac{4}{\pi ^2}-\frac{16 \log (2)}{\pi ^2}\biggr)+(1+z)(2+z)I_1+(1+z)(I_2+I_3),
\end{aligned}
\end{equation}
where 
\[
\begin{aligned}
I_1&=\frac{4}{\pi ^{3/2}}\int_{\gamma_u,\gamma_v} \frac{du}{2\pi i}\frac{dv}{2\pi i}\int_{\mathbb R_+^2} d a_3 da_4\frac{ \left(a_3+1\right)^{-\frac{v}{2}} \left(a_3+a_4\right){}^{\frac{v}{2}-\frac{3}{2}}  a_3^{\frac{w}{2}-1} \Gamma (u) \Gamma (v)\Gamma (w)\Gamma (\frac{3}{2}-\frac{v}{2})}{\left(a_4+z\right) \Gamma (\frac{u}{2}+1) \Gamma (w/2)},\\
I_2&=\frac{8}{\pi ^{3/2}}\int_{\gamma_u,\gamma_v} \frac{du}{2\pi i}\frac{dv}{2\pi i}\int_{\mathbb R_+^2} d a_3 da_4 \frac{\left(a_3+1\right)^{\frac{1}{2}-\frac{v}{2}} \left(a_3+a_4\right)^{\frac{v}{2}-1}  a_3^{\frac{w}{2}-\frac{3}{2}} \Gamma (u)  \Gamma (v)\Gamma (w)\Gamma (1-\frac{v}{2})}{\left(a_4+z\right) \Gamma (\frac{u}{2}+1) \Gamma (w/2)},\\
I_3&=-\int_{\mathbb R_+^2} d a_3 da_4\frac{2 }{\pi ^2 \sqrt{a_3} \sqrt{a_3+1} \left(a_3+a_4\right) \left(a_4+z\right)},
\end{aligned}
\]
and the contour here is
\[
\gamma_u=\frac{1}{3}+i \mathbb R,\quad 
\gamma_v=\frac{1}{4}+i \mathbb R.
\]

Note that we left two Feynman parameters to integrate later because if we directly integrate it, we will meet hypergeometric functions and it's very difficult to perform the inverse Mellin transform for it. 
However, we can introduce another Mellin transform for $1/(a_4+z)$ to avoid this difficulty so that integration of $a_3$ and $a_4$ only produces $\Gamma$ functions, and then the integration of the left $u$ and $v$ also becomes standard. Finally we get that
\begin{equation}
\begin{aligned}
I_1&=\pi ^{3/2}\int_{2/3-i\infty}^{2/3+i\infty}\frac{ds}{2\pi i}z^{-s} \left(\frac{2  \csc(\pi  s)^3 \sec (\pi  s)}{\Gamma (5/2-s) \Gamma (s-1)}-\frac{4  \csc(\pi  s)^3 \sec (\pi  s)}{(s-2) \Gamma (1-s) \Gamma (s-1/2)}\right),\\
I_2&=\pi ^{3/2}\int_{2/3-i\infty}^{2/3+i\infty}\frac{ds}{2\pi i}z^{-s} \left(\frac{8 \csc(\pi  s)^3 \sec (\pi  s)}{\Gamma (3-s) \Gamma (s-1/2)}-\frac{2\csc(\pi  s)^3 \sec (\pi  s)}{\Gamma (5/2-s) \Gamma (s)}\right),\\
I_3&=\pi ^{3/2}\int_{2/3-i\infty}^{2/3+i\infty}\frac{ds}{2\pi i}z^{-s} \frac{2\csc(\pi  s)^3 \sec (\pi  s)}{\Gamma (3/2-s) \Gamma (s)}.
\end{aligned}
\end{equation}

Therefore, after shifting contours and absorb the extra $z$ factors into inverse Mellin transformation, we have that
\begin{equation}
B_1=4\pi^{9/2}\biggl(\frac{2}{3}+2\sqrt{z}-\frac{4}{\pi^2}f(z)(1+z)\biggr),
\end{equation}
where the function $f$ is
\begin{equation}\label{wfz}
\begin{aligned}
f(z)&=\pi^{7/2}\int_{1/3-i\infty}^{1/3+i\infty}\frac{ds}{2\pi i}z^{-s} \frac{ \csc(\pi  s)^3 \sec (\pi  s)}{\Gamma (1-s) \Gamma (s+1/2)}\\
&=\int_0^\infty \frac{dw}{\sqrt{w (w+z)}}\frac{\log (w)}{w-1}\\
&=\frac{2}{\sqrt{1+z}}\left(\text{Li}_2(1-\tau)-\frac{1}{4} \log (\tau)^2+\log (\tau-1) \log (\tau)+\frac{\pi ^2}{2}\right)
\end{aligned}
\end{equation}
with $\tau=(\sqrt{1+z}+1)/(\sqrt{1+z}-1)$, and we use the convolution formula
\[
\int_\gamma \frac{ds}{2\pi i}z^{-s}\hat{f}(s)\hat{g}(s)=\int_0^\infty \frac{dw}{w}\, f(z/w)g(w)
\]
to get the first identity in eq.\eqref{wfz}, where $\hat f$ and $\hat g$ are Mellin transform of $f$ and $g$ respectively.

Finally, the contribution of all box diagram reads
\begin{equation}
\begin{aligned}
F_{\text{box}}&=\frac{1}{4!}\biggl(\frac{i}{2\sqrt{\pi}}\biggr)^{4}(3B_1z^{-1/4}+(z\leftrightarrow z^{-1}))\\
&=\frac{\pi^3}{32}\biggl(\frac{2}{3}z^{-1/4}+2z^{1/4}-\frac{4}{\pi^2}z^{-1/4}(1+z)f(z)\biggr)+(z\leftrightarrow 1/z).
\end{aligned}
\end{equation}


\subsection{Ladder diagram}\label{s:2.3}

The integrand of ladder diagram is 
\[
\begin{aligned}
L&=\begin{tikzpicture}[baseline={([yshift=-1.7ex]current bounding box.center)}]
\draw (-2,0) -- (-0.5,0);
\draw[fill=black] (-1,0) circle[radius=.5ex];
\draw[fill=white] (-1.5,0) circle[radius=.5ex];
\draw[fill=black] (-2,0) circle[radius=.5ex];
\draw[fill=white] (-0.5,0) circle[radius=.5ex];
\node at (-2,0.3) {$1$};
\node at (-1.5,0.3) {$2$};
\node at (-1,0.3) {$3$};
\node at (-0.5,0.3) {$4$};
\end{tikzpicture}\\
&=8 c^2
\frac{\epsilon_2 \epsilon_3}{D_{1,2} D_{2,3} D_{3,4} s_1  t_2 s_3 t_4}\\
&=-8 
\frac{\epsilon(\ell_2,1,2,3,4) \epsilon(\ell_3,1,2,3,4)}{\inp{\ell_1}{\ell_2}\inp{\ell_2}{\ell_3}\inp{\ell_3}{\ell_4} \inp{\ell_1}2\inp{\ell_1}4 \inp{\ell_2}1\inp{\ell_2}3\inp{\ell_3}2\inp{\ell_3}4\inp{\ell_4}1\inp{\ell_4}3}.
\end{aligned}
\]
Here we need to calculate $L_1$ and $L_3$. 

For $L_3$, we leave $\ell_3$ unintegrated, and the $\ell_1$ and $\ell_4$-integral is simply the one-loop scale triangle integral, so we get that
\[
L_3=-8\pi^3\frac{\epsilon(\ell,1,2,3,4)}{\inp{\ell}2\inp{\ell}4\sqrt{\inp{\ell}1\inp{\ell}3\inp13}}\int_{2}\frac{\epsilon(\ell_2,1,2,3,4)}{\inp{\ell_2}{\ell} \inp{\ell_2}1\inp{\ell_2}3\sqrt{\inp{\ell_2}2\inp{\ell_2}4\inp24}}.
\]
The $\ell_2$-integral is also a one-loop integral which can be integrated by introducing Feynman parameterization,
\[
\int_{2}\frac{\epsilon(\ell_2,1,2,3,4)}{\inp{\ell_2}{\ell} \inp{\ell_2}1\inp{\ell_2}3\sqrt{\inp{\ell_2}2\inp{\ell_2}4\inp24}}=\frac{1}{\sqrt{\pi}}\frac{\epsilon(\ell,1,2,3,4) }{\inp24\inp{\ell}1 \inp{\ell}3}
\int_{0}^\infty dt\,\frac{ \log(t z)}{\sqrt{t (1+t)}  (t z-1) },
\]
we introduce the inverse Mellin transformation 
\[
\frac{\log(tz)}{tz-1}=\pi^2\int_{1/3-i\infty}^{1/3+i\infty}\frac{ds}{2\pi i} (zt)^{-s} \csc(\pi  s)^2,
\]
and then perform the $t$ integral, we get that
\begin{equation}
\begin{aligned}
\int_{2}&\frac{\epsilon(\ell_2,1,2,3,4)}{\inp{\ell_2}{\ell} \inp{\ell_2}1\inp{\ell_2}3\sqrt{\inp{\ell_2}2\inp{\ell_2}4\inp24}}\\
&=
\pi^{3}\frac{ \epsilon(\ell,1,2,3,4)}{\inp24\inp{\ell}1 \inp{\ell}3}\int_{1/3-i\infty}^{1/3+i\infty}\frac{ds}{2\pi i}z^{-s} \frac{ \csc(\pi  s)^3 \sec (\pi  s)}{\Gamma (1-s) \Gamma (s+1/2)}.\\
\end{aligned}
\end{equation}
Therefore,
\[
\begin{aligned}
L_3&=-8 
\frac{\pi^{9/2}\epsilon(\ell,1,2,3,4)^2}{\inp{\ell}2\inp{\ell}4\sqrt{\inp{\ell}1\inp{\ell}3\inp13}}\frac{f(z)/\pi^2}{\inp24\inp{\ell}1 \inp{\ell}3}\\
&=-8
\frac{\pi^{9/2}\inp13\inp24}{\inp{\ell}2\inp{\ell}4\sqrt{\inp{\ell}1\inp{\ell}3\inp13}}\frac{(1+z)f(z)}{\pi^2}\\
&=-8\pi^{5}\frac{1}{\sqrt{\pi}}\biggl(\frac{(2\cdot 4)(1\cdot 3)}{(\ell\cdot 2)(\ell\cdot 4)(\ell\cdot 1)(\ell\cdot 3)}\biggr)^{3/4}z^{-1/4}\frac{(1+z)f(z)}{\pi^2},
\end{aligned}
\] 
where $f(z)$ is a weight-$2$ polylogarithm introduced in eq.\eqref{wfz}.

For $L_1$, we can first integrate $\ell_2,\ell_4$, and get a similar integral we meet in the integration of the box diagram, 
\[
\int_{\ell_2,\ell_4}L=\frac{P_L}{H}+Q_L,
\]
where $P_L,Q_L$ are similarly rational functions of scale products and $\epsilon$'s and 
\[
H=\sqrt{(1\cdot 3)(\ell_1\cdot \ell_3)}+\sqrt{(\ell_1\cdot 3)(\ell_3\cdot 1)}+\sqrt{(\ell_1\cdot 1)(\ell_3\cdot 3)}.
\]
Therefore, we follow the same method in the integration of the box diagram, by introducing Mellin transformation for $1/H$, and we get that
\begin{equation}
L_1=-8\pi^{5}\frac{1}{\sqrt{\pi}}\biggl(\frac{(2\cdot 4)(1\cdot 3)}{(\ell\cdot 2)(\ell\cdot 4)(\ell\cdot 1)(\ell\cdot 3)}\biggr)^{3/4}z^{-1/4}\frac{\log(z/4)^2+2\pi^2}{2\pi ^2}.
\end{equation}

Therefore, 
\begin{equation}
\begin{aligned}
F_{\text{ladder}}=24\,\,\begin{tikzpicture}[baseline={([yshift=-.5ex]current bounding box.center)}]
\draw (-2,0) -- (-0.5,0);
\draw[fill=white] (-1,0) circle[radius=.5ex];
\draw[fill=black] (-1.5,0) circle[radius=.5ex];
\draw[fill=white] (-2,0) circle[radius=.5ex];
\draw[fill=black] (-0.5,0) circle[radius=.5ex];
\end{tikzpicture}&=\frac{1}{4!}\biggl(\frac{i}{2\sqrt{\pi}}\biggr)^{4}(6L_1+6L_3+(z\leftrightarrow z^{-1}))\\
&=-\frac{\pi^3}{8}z^{-1/4}\biggl(1+\frac{\log(z/4)^2}{2\pi^2}+\frac{(1+z)f(z)}{\pi^2}\biggr)+(z\leftrightarrow z^{-1}).
\end{aligned}
\end{equation}


\section{Cusp anomalous dimension from integrated results}\label{sec:3}

In this section, we extract the cusp anomalous dimension from the functions obtained from the integration of  negative geometry. This can be done by integrating out the final loop variable: $\mathcal{W}_{L}$ diverges and the cusp anomalous dimension is encoded in the coefficient of $\epsilon^{-2}$ when the integral is done in $D=3- 2 \epsilon$. 

In \cite{Gromov:2008qe}, it is proposed from the all loop $\text{AdS}_4/\text{CFT}_3$ Bethe ansatz that the ABJM cusp anomalous dimension can be obtained from the $\mathcal N=4$ sYM cusp anomalous dimension by simple replacing the interpolating function $h^{\text{sYM}}$  to $h^{\mathrm{ABJM}}$:
\[
\Gamma_{\text {cusp }}^{\mathrm{ABJM}}=\left.\frac{1}{4} \Gamma_{\text {cusp }}^{\text{sYM}}\right|_{h^{\text{sYM}} \rightarrow h^{\mathrm{ABJM}}},
\]
where the interpolating function for $\mathcal N=4$ sYM is the square root of the coupling
\begin{equation}
h^{\text{sYM}}(\lambda)=\frac{\sqrt{\lambda}}{4 \pi},
\end{equation}
while the interpolating function for ABJM is a complicated function in $\lambda$ ~\cite{Gromov:2014eha}
\begin{equation}
\lambda=\frac{\sinh (2 \pi h^{\mathrm{ABJM}})}{2 \pi}{ }_3 F_2\left(\frac{1}{2}, \frac{1}{2}, \frac{1}{2} ; 1, \frac{3}{2} ;-\sinh (2 \pi h^{\mathrm{ABJM}})^2\right) .
\end{equation}
Therefore, from the weak coupling expansion of $\Gamma_{\text {cusp }}^{\text{sYM}}$ \cite{Beisert:2006ez}
\begin{equation}
\Gamma_{\text {cusp }}^{\text{sYM}}(h)=4 h^2-\frac{4}{3} \pi^2 h^4+\frac{44}{45} \pi^4 h^6+\cdots,
\end{equation}
we get that
\begin{equation}\label{Gcusp}
\Gamma_{\text {cusp }}^{\mathrm{ABJM}}(\lambda)=\lambda^2-\pi^2 \lambda^4+\frac{49 \pi^4}{30} \lambda^6+\cdots,
\end{equation}
where the leading order at $L=2$ is checked in \cite{Griguolo:2012iq,He:2023exb,Henn:2023pkc}. Here we are going to check $\Gamma_{\text {cusp }}^{\mathrm{ABJM}}(\lambda)^{(4)}=-\pi^2$.

To evaluate the last loop integration and extract $\epsilon^{-2}$ divergence, one can expand $F_{L-1}(z)$ in $z$ around $z=0$. This series expansion has logarithmic divergence in general: 
\begin{equation}
    F_{L-1}(z)=\sum_{p,q}c^{(L-1)}_{p,q}z^p\log(z)^q,
\end{equation}
where $q$ is a non-negative integer, and $p$ can be any rational number. Notice that 
\[
z^p\log(z)^q=\frac{\partial^q}{\partial p^q}z^p,
\]
so we only need to calculate the integral for $z^p$. As shown in~\cite{He:2023exb, Henn:2023pkc}, the $\epsilon^{-2}$ divergence of the $z^p$ integral in $D=3-2\epsilon$ is simply as
\begin{equation}\label{pgamma}
    \int_{\ell_1} \left(\frac{(1\cdot 3)(2\cdot 4)}{(\ell_1\cdot 1)(\ell_1\cdot 2)(\ell_1\cdot 3)(\ell_1\cdot 4)}\right)^{\frac{3}{4}} z^p
    = \frac{2}{\sqrt{\pi}} \frac{1}{\Gamma\left(3/4-p\right) \Gamma\left(3/4+p\right)}\epsilon^{-2}+O(\epsilon^{-1}).
\end{equation}

Then, we can replace 
\[
z^{\pm 1/4}\to \frac{2}{\pi},\quad z^{\pm 1/4}\log(z)\to \mp\frac{4 \log (2)}{\pi },\quad 
z^{\pm 1/4}\log(z)^2\to -\frac{4 \pi }{3}+\frac{8 \log(2) ^2}{\pi },
\]
and 
\[
\begin{aligned}
\frac{1}{\pi^2}z^{-1/4}(1+z)&f(z)\\
&=\pi ^{3/2}\int_{1/3-i\infty}^{1/3+i\infty}\frac{ds}{2\pi i}(z^{-1/4-s}+z^{3/4-s}) \frac{ \csc(\pi  s)^3 \sec (\pi  s)}{\Gamma (1-s) \Gamma \left(s+\frac{1}{2}\right)}\\
&\to 
2\pi\int_{1/3-i\infty}^{1/3+i\infty}\frac{ds}{2\pi i}\biggl(\frac{1}{\Gamma (\frac{1}{2}-s) \Gamma (s+1)}+\frac{1}{\Gamma(\frac{3}{2}-s) \Gamma (s)}\biggr) \frac{\csc(\pi  s)^3 \sec (\pi  s)}{\Gamma (1-s) \Gamma \left(s+\frac{1}{2}\right)}\\
&=\frac{2}{\pi}\int_{1/3-i\infty}^{1/3+i\infty}\frac{ds}{2\pi i}\frac{\csc (\pi  s)^2}{(1-2 s) s}\\
&=\frac{4}{3\pi}
\end{aligned}
\]
in $F_{\text{star}}$, $F_{\text{ladder}}$ and $F_{\text{box}}$ to get the corresponding contribution to the $4$-loop cusp anomalous dimension $\Gamma_{\text{cusp}}^{(4)}$:
\begin{align}
F_{\text{star}}&=-\frac{\pi^{3}}{12}(z^{-1/4}+z^{1/4})\to -\frac{\pi^2}{3},\\
F_{\text{ladder}}&=-\frac{\pi^3}{8}z^{-1/4}\biggl(1+\frac{\log(z/4)^2}{2\pi^2}+\frac{(1+z)f(z)}{\pi^2}\biggr)+(z\leftrightarrow z^{-1})\nonumber\\
&\to 
2\times\frac{\pi^3}{8}\biggl(-\frac{2}{\pi}+\frac{2}{3\pi}-\frac{4}{3\pi}\biggr)
=-\frac{2\pi^2}{3}.
\end{align}
but the box does not contribute the $\Gamma_{\text{cusp}}$
\begin{align}
F_{\text{box}}&=\frac{\pi^3}{32}\biggl(\frac{7}{3}(z^{-1/4}+z^{1/4})-\frac{4}{\pi^2}z^{-1/4}(1+z)f(z)-\frac{4}{\pi^2}z^{1/4}(1+1/z)f(1/z)\biggr)\nonumber\\
&\to 2\times \frac{\pi^3}{32}\biggl(\frac{8}{3}\frac2{\pi}-4\frac{4}{3\pi}\biggr)=0.
\end{align}
Therefore, $4$-loop cusp anomalous dimension is
\begin{equation}
\Gamma_{\text{cusp}}^{(4)}=-\frac{\pi^2}{3}-\frac{2\pi^2}{3}=-\pi^2,
\end{equation}
which agrees with eq.\eqref{Gcusp} at $L=4$.

Note that the box diagram does not contribute to $\Gamma_{\text{cusp}}$, which means that if we integrate all loop momenta in the box diagram, we have that
\begin{equation}
\int\prod_{i=1}^4d^3\ell_i\begin{tikzpicture}[baseline={([yshift=-.5ex]current bounding box.center)},scale=0.75]
\draw (-2,-1.5) -- (-1,-1.5) -- (-1,-0.5) -- (-2,-0.5) -- cycle;
\draw[fill=white] (-1,-1.5) circle[radius=.5ex];
\draw[fill=black] (-1,-0.5) circle[radius=.5ex];
\draw[fill=white] (-2,-0.5) circle[radius=.5ex];
\draw[fill=black] (-2,-1.5) circle[radius=.5ex];
\node at (-2.2,-1.7) {$1$};
\node at (-0.8,-1.7) {$2$};
\node at (-0.8,-0.3) {$3$};
\node at (-2.2,-0.3) {$4$};
\end{tikzpicture}= O(\epsilon^{-1}).
\end{equation}
If we go back to eq.\eqref{l4neg2}, the integrand of the box diagram can be decomposed into three parts with $0$, $2$ and $4$ $\epsilon_i$'s on its numerator as 
$B=N_0+N_2+N_4$, and we find that
\[
\int\prod_{i=1}^4d^3\ell_iN_4=-\int\prod_{i=1}^4d^3\ell_iN_2\propto \log (2)\epsilon^{-2}+O(\epsilon^{-1}),\quad \int\prod_{i=1}^4d^3\ell_iN_0=O(\epsilon^{-1}).
\]
Therefore, it is so non-trivial that the construction of the integrand of the box diagram weakens the IR divergence. It will be interesting to see this directly from the integrand. The IR divergence of the Feynman integral in $\mathcal N=4$ sYM appears when one of the loop momentum goes to a collinear region \cite{Arkani-Hamed:2013kca,Arkani-Hamed:2021iya} 
\begin{equation}
\ell_i\to ax_k+(1-a)x_{k+1}+\eta k_i
\end{equation}
for $k=1,\dots,4$, where $a$ is a parameter, $k_i$ is a reference momentum and $\eta$ is a small positive number. In \cite{Arkani-Hamed:2021iya}, it is argued that if one loop momentum is in a collinear region, then the condition eq.\eqref{ineq} requires that all loop momentua are in the same collinear region, which weakens the divergence of a $L$-loop integral to a one-loop divergence $O(\epsilon^{-2})$. Since there are further constrains on negative geometries of ABJM theory \cite{He:2022cup}, it would be possible that the divergence could be further weakened for some negative geometries. We leave this for further work.

\section{Higher loop ladders}
\label{sec:4}

In this section, we consider a special kind of all-loop negative geometries corresponding to ladder diagrams.

Integrands of ladder diagrams have a recursive structure \cite{He:2022cup}
\[
\begin{tikzpicture}[baseline={([yshift=-1.7ex]current bounding box.center)}]
\draw (-2,0) -- (2.5,0);
\draw[fill=black] (-1,0) circle[radius=.5ex];
\draw[fill=white] (-1.5,0) circle[radius=.5ex];
\draw[fill=black] (-2,0) circle[radius=.5ex];
\draw[fill=white] (-0.5,0) circle[radius=.5ex];
\draw[fill=black] (0,0) circle[radius=.5ex];
\draw[fill=white] (2.5,0) circle[radius=.5ex];
\draw[fill=black] (2,0) circle[radius=.5ex];
\node at (-2,0.3) {$1$};
\node at (-1.5,0.3) {$2$};
\node at (-1,0.3) {$3$};
\node at (-0.5,0.3) {$4$};
\node at (0,0.3) {$5$};
\node at (2.7,0.331) {$n{+}1$};
\node at (2,0.3) {$n$};
\node at (1,0.2) {$\cdots\cdots$};
\end{tikzpicture}\, = \,
\begin{tikzpicture}[baseline={([yshift=-1.7ex]current bounding box.center)}]
\draw (-2,0) -- (2,0);
\draw[fill=black] (-1,0) circle[radius=.5ex];
\draw[fill=white] (-1.5,0) circle[radius=.5ex];
\draw[fill=black] (-2,0) circle[radius=.5ex];
\draw[fill=white] (-0.5,0) circle[radius=.5ex];
\draw[fill=black] (0,0) circle[radius=.5ex];
\draw[fill=black] (2,0) circle[radius=.5ex];
\node at (-2,0.3) {$1$};
\node at (-1.5,0.3) {$2$};
\node at (-1,0.3) {$3$};
\node at (-0.5,0.3) {$4$};
\node at (0,0.3) {$5$};
\node at (2,0.3) {$n$};
\node at (1,0.2) {$\cdots\cdots$};
\end{tikzpicture}\times \frac{2\epsilon_{n}}{D_{n,n+1}t_{n+1}}
\]
and the initial ladder is 
\[
\begin{tikzpicture}[baseline={([yshift=-1.7ex]current bounding box.center)}]
\draw (-1.6,0) -- (-2,0);
\draw[fill=black] (-2,0) circle[radius=.5ex];
\draw[fill=white] (-1.5,0) circle[radius=.5ex];
\node at (-2,0.3) {$1$};
\node at (-1.5,0.3) {$2$};
\end{tikzpicture}=\frac{2c^2}{s_1t_2D_{12}}.
\]
For these two ladder $L_{n+1}$ and $L_n$, if we already have the integral of $L_n$ with only $\ell_n$ unintegrated
\[
\int_{1,\dots,n-1}L_{n}=\alpha_{\text{black}}(\ell_n)\tilde L_{n}(z_n),
\]
where $\alpha_{\text{black}}(\ell_n)$ is a non-DCI prefactor which makes $\tilde L_{n}$ be the function in $z_n$. The prefactor is universal for nodes with the same color, and $\alpha_{\text{white}}$ is given by interchanging dual points $X_1\leftrightarrow X_2,X_3 \leftrightarrow X_4$ in $\alpha_{\text{black}}$.
Following the prefactor eq.\eqref{eq:pref} used in the calculating of box diagram, we choose 
\begin{equation}
\alpha_{\text{black}}(\ell)=\frac{(2\cdot 4)\sqrt{(1\cdot 3)}}{(\ell\cdot 2)(\ell\cdot 4)\sqrt{(\ell\cdot 1)(\ell\cdot 3)}},\quad 
\alpha_{\text{white}}(\ell)=\frac{(1\cdot 3)\sqrt{(2\cdot 4)}}{(\ell\cdot 1)(\ell\cdot 3)\sqrt{(\ell\cdot 2)(\ell\cdot 4)}},
\end{equation}
so $\tilde L_2=2\pi^{3/2}$.
Therefore, the recursion relation becomes
\begin{equation}\label{eq:rec}
\alpha_{\text{white}}(\ell_{n+1})\tilde L_{n+1}(z_{n+1})=\int_{1,\dots,n}L_{n+1}=\int_{n} \frac{2\epsilon_{n}}{D_{n,n+1}t_{n+1}}\alpha_{\text{black}}(\ell_n)\tilde L_{n}(z_n)
\end{equation}
and the similar formula for ladders with the inverse color.

We can see eq.\eqref{eq:rec} as an integral transformation of  $\hat L_{n}(z_n)$ in $\ell_{n}$-space with the kernel 
\[
\frac{2\epsilon_{n}}{D_{n,n+1}t_{n+1}}\alpha_{\text{black}}(\ell_n)=\frac{2}{(\ell_{n+1}\cdot 1)(\ell_{n+1}\cdot 3)}\frac{\epsilon(1,2,3,4,\ell_n)}{(\ell_n\cdot\ell_{n+1})(\ell_n\cdot 2)(\ell_n\cdot 4)\sqrt{(\ell_n\cdot 1)(\ell_n\cdot 3)}}.
\]
To perform the loop integral, we can further choose a Mellin representation $\tilde L_{n}(z_n)$ as 
\[
\tilde L_{n}(z_n)=\int_{\gamma}\frac{ds}{2\pi i} \hat L_{n}(s)z_n^{-s},
\]
then the recursion relation becomes a new integral transformation in the Mellin $s$-space with the kernel 
\[
K(s)=\int_n\frac{\epsilon(1,2,3,4,\ell_n)}{(\ell_n\cdot\ell_{n+1})(\ell_n\cdot 2)(\ell_n\cdot 4)\sqrt{(\ell_n\cdot 1)(\ell_n\cdot 3)}}z_n^{-s},
\]
and we can first integrate $\ell_n$ in this kernel.
From the definition of $z_n$ eq.\eqref{defz}, $K(s)$ is a one-loop integral of the form eq.\eqref{A1}, and its Mellin 
representation is given in the Appendix:
\[
K(s)=\int_{\gamma} \frac{dt}{2\pi i}z_{n+1}^{-t}\frac{\sqrt{\pi } \csc(\pi  t) ^2 \Gamma (s-t) \Gamma (-s+t+\frac{1}{2})}{\Gamma (\frac{1}{2}-s)^2 \Gamma (s+1)^2}\frac{(1\cdot 3)(2\cdot 4)\epsilon(1,2,3,4,\ell_{n+1})}{(\ell_{n+1}\cdot 1) (\ell_{n+1}\cdot 3)(\ell_{n+1}\cdot 2) (\ell_{n+1}\cdot 4)}.
\]
Therefore, we get the recursion relation in Mellin space
\begin{equation}\label{mrec}
\frac{1}{\sqrt{1+z^{-1}}}\tilde L_{n+1}(z)=
2\sqrt{\pi}\int_{\gamma_s,\gamma_t}\frac{dsdt}{(2\pi i)^2} z^{-t}\frac{ \csc(\pi  t)^2 \Gamma (-s+t+1/2) \Gamma (-t+s)}{\Gamma (1+s)^2 \Gamma (1/2-s)^2 }\hat L_{n}(s),
\end{equation}
and the contour is choosen such that
\[
\operatorname{Re}[-s+t+1/2]>0,\quad \operatorname{Re}[-t+s]>0, \quad 
-1<\operatorname{Re}[t]<0.
\]
Similarly the further recursion to $\tilde L_{n+2}(z)$ is in the invert color, so it is given by replacing $z\to z^{-1}$ (and correspondingly $s,t\to -s,-t$) in eq.\eqref{mrec}:
\begin{equation}\label{mrec2}
\frac{1}{\sqrt{1+z}}\tilde L_{n+2}(z)=2\sqrt{\pi}
\int_{\gamma_s,\gamma_t}\frac{dsdt}{(2\pi i)^2} z^{-t}\frac{ \csc(\pi  t)^2 \Gamma (s-t+1/2) \Gamma (t-s)}{\Gamma (1-s)^2 \Gamma (1/2+s)^2 }\hat L_{n+1}(s).
\end{equation}

As an example, we first consider the following integral
\[
\begin{tikzpicture}[baseline={([yshift=-1.7ex]current bounding box.center)}]
\draw (-1.6,0) -- (-2,0);
\draw[fill=white] (-2,0) circle[radius=.5ex];
\draw[fill=black] (-1.5,0) circle[radius=.5ex];
\node at (-2,0.3) {$1$};
\node at (-1.5,0.3) {$2$};
\end{tikzpicture}=\frac{2c^2}{t_1s_2D_{12}},
\quad 
\begin{tikzpicture}[baseline={([yshift=-1.7ex]current bounding box.center)}]
\draw (-2,0) -- (-1,0);
\draw[fill=white] (-1,0) circle[radius=.5ex];
\draw[fill=black] (-1.5,0) circle[radius=.5ex];
\draw[fill=white] (-2,0) circle[radius=.5ex];
\node at (-2,0.3) {$1$};
\node at (-1.5,0.3) {$2$};
\node at (-1,0.3) {$3$};
\end{tikzpicture}=\frac{4c^2 \epsilon_2}{t_1s_2t_3D_{12}D_{23}}.
\]
It is easy to perform $L_2$ by using the one-loop scale triangle:
\begin{equation}
    \tilde L_2(z)=2\pi^{3/2}\quad \text{or}\quad \hat L_2(s)=2\pi^{3/2}2\pi i\,\delta(s),
\end{equation}
therefore from eq.\eqref{mrec}
\[
\frac{1}{\sqrt{1+z^{-1}}}\tilde L_3(z)=
2\pi^3\int_{-1/4-i\infty}^{-1/4+i\infty}\frac{dt}{2\pi i} z^{-t} \frac{ \csc (\pi  t)^3 \sec (\pi  t)}{\Gamma (1/2-t)\Gamma (1+t)}=\frac{2}{\sqrt{\pi}}f(1/z),
\]
where the function $f(z)$ is defined in eq.\eqref{wfz}, so $\tilde L_3$ is a pure weight-2 polylogarithm
\begin{equation}
\tilde L_3(z)=\frac{4}{\sqrt \pi}\left(\text{Li}_2(1-x)-\frac{1}{4} \log (x)^2+\log (x-1) \log (x)+\frac{\pi ^2}{2}\right)
\end{equation}
with $x=(\sqrt{1+z^{-1}}+1)/(\sqrt{1+z^{-1}}-1)$, which is first evaluated in \cite{Henn:2023pkc,He:2023exb}, and its symbol alphabet is $\{x,1-x\}$.

Similarly, we can calculate $\tilde L_5$ from $\tilde L_4(z)$ by eq.\eqref{mrec}. The Mellin transformation of $\log(z/4)^2+2\pi^2$ is 
\[
\hat L_4(s)\propto 2\pi^2+(\partial_s-\log(4))^2 2\pi i\,\delta(s),
\]
so after a little simplification and using the convolution formula
\[
\int_\gamma \frac{ds}{2\pi i}z^{-s}\hat{f}(s)\hat{g}(s)=\int_0^\infty \frac{dw}{w}\, f(z/w)g(w),
\]
we get that 
\begin{equation}\label{5loopladder}
\frac{1}{\sqrt{1+z^{-1}}}\tilde L_5(z)\propto \int_0^\infty \frac{z\,dw}{ \sqrt{w(w+1)} (w-z)}\log \left(\frac{z}{w}\right)\left(\log(4w)^2+\frac{2 \pi ^2}{3}\right).
\end{equation}
By replacing $w=(1-y)^2/(4y)$ and $z=(1-x)^2/(4x)$, the integral becomes
\[
\tilde L_5(z)\propto 
\int_{0<y<1} d\log\left(\frac{1-x y}{y-x}\right)\log \left(\frac{(x-1)^2 y}{x (y-1)^2}\right)\left(\log\biggl(\frac{(1-y)^2}{y}\biggr)^2+\frac{2 \pi ^2}{3}\right),
\]
which is a standard iterated integral of polylogarithms, and it has the same symbol alphabet as $\tilde L_3(z)$. We list the function of this integral in the Appendix.

The calculation of $\tilde L_4(z)$ from recursion relation is quite tricky. 
We want to derive it from $\tilde L_2(z) \propto 1$ by first apply eq.\eqref{mrec} and then eq.\eqref{mrec2}. However, 
we already know that 
\[
\tilde L_4(z)\propto \log(z/4)^2+2\pi^2
\]
is a polynomial of $\log(z/4)$ from the concrete calculation in section \ref{s:2.3}. 
We find it is only possible to first take the residue around $-s+t+1/2=0$ in eq.\eqref{mrec2}, and $z^{-t}\to z^{-s+1/2}$ which produce an extra $\sqrt{z}$ to cancel the a $\sqrt{z}$ from $\sqrt{1+1/z}$ in eq.\eqref{mrec}, and then we need to take the residue around $s=0$ in eq.\eqref{mrec2} since other pole will produce a power of $z$.
After taking above residues and removing irrelevant factors, we re-obtain that
\[
\tilde L_4(z)\propto \oint_{t=0} \frac{dt}{2 \pi i}z^{-t} \frac{4 \sqrt{\pi }  \csc (2 \pi  t)^2}{t\,\Gamma (1-t) \Gamma(t+1/2)}\propto \log(z/4)^2+2\pi^2
\]
through this rough reasoning. 

Similarly, for the recursion from even loop ladder $\tilde L_{2n}(z)$ to $\tilde L_{2n+2}(z)$, we conjecture that if there is a term 
\[
\log(z)^k=\left.\frac{\partial^k}{\partial p^k}z^p\right|_{p=0}
\]
in the $\tilde L_{2n}(z)$, it will produce a term
\begin{equation}
T[\log(z)^k]= -4 \pi ^{3/2}\oint_{t=0} \frac{dt}{2 \pi i} z^{-t}\left.\frac{\partial^k}{\partial p^k}\frac{ \csc(2 \pi  t)^2 \Gamma (p+t+1/2) \Gamma (-t-p)}{\Gamma (1-p)^2 \Gamma (p+1/2)^2 \Gamma (1/2+t)^2 \Gamma (1-t)^2}\right|_{p=0}
\end{equation}
in the $\tilde L_{2n+2}(z)$. Since $\tilde L_{2n}(z)$ would be a function in $\log(z/4)$, it is also convenient to use the variant 
\begin{equation}
T[\log(z/4)^k]= -4 \pi ^{3/2}\oint_{t=0} \frac{dt}{2 \pi i} z^{-t}\left.\frac{\partial^k}{\partial p^k}\frac{ 4^{-p}\csc(2 \pi  t)^2 \Gamma (p+t+1/2) \Gamma (-t-p)}{\Gamma (1-p)^2 \Gamma (p+1/2)^2 \Gamma (1/2+t)^2 \Gamma (1-t)^2}\right|_{p=0}.
\end{equation}
The integral has a closed formula:
\begin{equation}\label{Tlog}
T[\log(z/4)^k]=\frac{\log (z/4)^{k+2}}{\pi ^2 (k+1) (k+2)}+B_k \log (z/4)+C_k
\end{equation}
with
\begin{equation}\label{Bdef}
B_k=\frac{1}{(k+1) (k+2)}\left.\frac{\partial^{k+2}}{\partial p^{k+2}}\frac{4^{-p}}{\pi ^{3/2} \Gamma (-p) \Gamma (p+1/2)}\right|_{p=0},
\end{equation}
\begin{equation}\label{Cdef}
C_k=-\frac{1}{(k+1) (k+2)}\left.\frac{\partial^{k+2}}{\partial p^{k+2}}\frac{4^{-p} (-\psi(-p)+\psi (p+1/2)+\log (4))}{ \pi ^{3/2} \Gamma (-p) \Gamma (p+1/2)}\right|_{p=0},
\end{equation}
where $\psi(z)=\frac{d}{dz}\log(\Gamma(z))$ is the digamma function. 

Therefore, from $\tilde L_4\propto \log(z/4)^2+2\pi^2$, our conjecture predicts that 
\begin{align}
\tilde L_6=T[\tilde L_4]\propto & -4 \zeta_3 \log \left(\frac{z}{4}\right)+\frac{1}{12} \log \left(\frac{z}{4}\right)^4+\pi ^2 \log \left(\frac{z}{4}\right)^2+\frac{17 \pi ^4}{9},\\
\tilde L_8=T[\tilde L_6]\propto &-\frac{2}{3} \zeta_3 \log \left(\frac z4\right)^3-4 \pi ^2 \zeta_3 \log \left(\frac z4\right)-12 \zeta_5 \log \left(\frac z4\right)+4 \zeta_3^2+\nonumber\\
&\frac{1}{360}\log \left(\frac z4\right)^6+\frac{1}{12} \pi ^2 \log \left(\frac z4\right)^4+\frac{17}{18} \pi ^4 \log \left(\frac z4\right)^2+\frac{49 \pi ^6}{27}
\end{align}
and so on. Note that since
\[
\pi^2(z\partial_z)^2 T[\log(z/4)^k]=\log(z/4)^k,
\]
we have a differential equation to relate $\tilde L_{2n}$ with $\tilde L_{2n+2}$:
\begin{equation}\label{ladderde}
\pi^2(z\partial_z)^2 \tilde L_{2n+2}=\tilde L_{2n}.
\end{equation}

The computation of cusp anomalous dimension will also provide a non-trivial check for our conjecture. For example, we can compute the $8$-loop ladder with the node $5$ unintegrated (suppose that the node $5$ is black), then it becomes the product of two ladders with length $4$ and $5$, and 
\begin{equation}
\int_{1,2,3,4}\int_{6,7,8}L_8 \propto (z^{1/4}+z^{-1/4}) \tilde L_4 \frac{\tilde L_5}{\sqrt{1+z^{-1}}},
\end{equation}
and its contribution to cusp anomalous dimension should equal to the contribution from $\tilde L_8$ since they are the same leading divergence of $\epsilon^{-2}$ of 
\[
\int_{1,2,3,4,5,6,7,8}L_8.
\]

Despite the correctness of conjectures, it is interesting to resum these ladders. According to the differential equation eq.\eqref{ladderde} for $n\geq 1$
\[
\pi^2(z\partial_z)^2 \tilde L_{2n+2}=\tilde L_{2n},
\]
we can define $K(\lambda,z)=\sum_{n=1}^\infty(\pi \lambda)^{2n}\tilde L_{2n}$, then there is a differential equation for resummed $K(\lambda,z)$
\[
(z\partial_z)^2K(\lambda,z)=\lambda^2K(\lambda,z),
\]
the solution is 
\begin{equation}
K(\lambda,z)=c_1(\lambda) \exp (\lambda \log (\frac z4))+c_2(\lambda) \exp (-\lambda \log (\frac z4))
\end{equation}
depending on two unknown functions $c_1(\lambda)$ and $c_2(\lambda)$. Conversely, we can also use the integral equation $T[\tilde L_{2n}]=\tilde L_{2n+2}$ to the resummed $K(\lambda,z)$ and get that 
\begin{equation}\label{ieT}
\pi^2\lambda^2T[K(\lambda,z)]=\sum_{n=1}^\infty (\pi\lambda)^{2n+2} \tilde L_{2n+2}=K(\lambda,z)-\pi^2\lambda^2 \tilde L_{2}.
\end{equation}
Therefore, by using eq.\eqref{Tlog}, the coefficients of $\log(z/4)^0$ and $\log(z/4)^1$ of eq.\eqref{ieT} give
\[
-\lambda^2\tilde L_2+\frac{c_1 + c_2}{\pi^2}=\sum_{k=0}^\infty\frac{c_1 +(-1)^k c_2}{k!}\lambda^{k+2}C_k,
\]
and 
\[
\frac{c_1 - c_2}{\pi^2}\lambda=\sum_{k=0}^\infty\frac{c_1 +(-1)^k c_2}{k!}\lambda^{k+2}B_k,
\]
and the other coefficients do not contribute more constrains. According to the definition of $B_k$ eq.\eqref{Bdef} and $C_k$ eq.\eqref{Cdef}, we get that 
\[
\left\{
\begin{aligned}
c_1 F(\lambda)+c_2 F(-\lambda)&=\lambda^2\tilde L_2\\
c_1 G(\lambda)+c_2 G(-\lambda)&=0
\end{aligned}
\right.
\]
where
\begin{equation}
F(p):=\frac{4^{-p} (-\psi(-p)+\psi (p+1/2)+\log (4))}{ \pi ^{3/2} \Gamma (-p) \Gamma (p+1/2)},\quad G(p):=\frac{4^{-p}}{\pi ^{3/2} \Gamma (-p) \Gamma (p+1/2)},
\end{equation}
and the solution
\begin{equation}
\begin{aligned}
c_1 &=\frac{\lambda ^2\tilde L_2 G(-\lambda)}{F(\lambda) G(-\lambda)-F(-\lambda) G(\lambda)}=-\frac{2 \pi ^3 \lambda ^2 \tilde L_2 G(-\lambda)}{\sin (2\pi \lambda )+2 \pi\lambda  \cos (2 \pi\lambda )},\\
c_2&= -\frac{\lambda ^2 \tilde L_2 G(\lambda)}{ F(\lambda) G(-\lambda)-F(-\lambda) G(\lambda)}=\frac{2 \pi ^3 \lambda ^2 \tilde L_2 G(\lambda)}{\sin (2 \pi\lambda )+2 \pi\lambda  \cos (2 \pi\lambda )},
\end{aligned}
\end{equation}
so we get the resummed ladder as
\begin{equation}
K(\lambda,z)=\frac{2 \pi ^3 \lambda ^2 \tilde L_2 }{\sin (2 \pi\lambda )+2 \pi\lambda  \cos (2 \pi\lambda )} \biggl(G(\lambda)\exp (-\lambda \log (\frac z4))- G(-\lambda)\exp (\lambda \log (\frac z4))\biggr).
\end{equation}

To compute its contribution to $\Gamma_{\text {cusp }}$, we can simply apply the replacement from eq.\eqref{pgamma}
\[
\log(z/4)^k\to \sqrt{\pi}\biggl.\frac{\partial^k}{\partial p^k}\biggr|_{p=-1/4}\frac{4^{-p-1/4}}{\Gamma\left(3/4-p\right) \Gamma\left(3/4+p\right)}=\sqrt{\pi}\biggl.\frac{\partial^k}{\partial p^k}\biggr|_{p=0}\frac{4^{-p}}{\Gamma\left(1/2+p\right) \Gamma\left(1-p\right)},
\]
so the contribution is 
\begin{equation}\label{Gladder}
\begin{aligned}
\Gamma_{\text{ladder}}(\lambda):=&\,\frac{2 \pi ^3 \lambda ^2  \sqrt{\pi}}{\sin (2 \pi\lambda )+2 \pi\lambda  \cos (2 \pi\lambda )} \biggl(G(\lambda)H(-\lambda)- G(-\lambda)H(\lambda) \biggr)\\
=\,&
\frac{2 \lambda ^2}{2 \pi  \lambda  \cot (2 \pi  \lambda )+1}\\
=\,&
\lambda ^2+\frac{2 \pi ^2 \lambda ^4}{3}+\frac{28 \pi ^4 \lambda ^6}{45}+\frac{568 \pi ^6 \lambda ^8}{945}+\frac{8272 \pi ^8 \lambda ^{10}}{14175}+\frac{9824 \pi ^{10} \lambda ^{12}}{17325}+O(\lambda ^{14})
\end{aligned}
\end{equation}
where 
\[
H(p):=\frac{4^{-p}}{\Gamma\left(1/2+p\right) \Gamma\left(1-p\right)}.
\]
Therefore, although there are zeta values like $\zeta_3,\zeta_5$ in $\tilde L_{2n+2}$, they do not contribute to the cusp anomalous dimension.
Note that $\Gamma_{\text{ladder}}(\lambda)$ have infinite poles along the real $\lambda$-axis, and its weak coupling expansion only provides positive series coefficient, while the other diagrams will convert the sign  in the whole cusp anomalous dimension eq.\eqref{Gcusp}, which also reduces the exponential divergence of eq.\eqref{Gladder} at the strong coupling.

\section{Discussions}

In this note, we have integrated the ABJM negative geometries at $L=4$ and ladder-type negative geometries at all loop by recursion relation, both based on introducing suitable Mellin transformation.

One of most interesting result at $L=4$ is that the box diagram
\[
\int\prod_{i=1}^4d^3\ell_i\begin{tikzpicture}[baseline={([yshift=-.5ex]current bounding box.center)},scale=0.75]
\draw (-2,-1.5) -- (-1,-1.5) -- (-1,-0.5) -- (-2,-0.5) -- cycle;
\draw[fill=white] (-1,-1.5) circle[radius=.5ex];
\draw[fill=black] (-1,-0.5) circle[radius=.5ex];
\draw[fill=white] (-2,-0.5) circle[radius=.5ex];
\draw[fill=black] (-2,-1.5) circle[radius=.5ex];
\node at (-2.2,-1.7) {$1$};
\node at (-0.8,-1.7) {$2$};
\node at (-0.8,-0.3) {$3$};
\node at (-2.2,-0.3) {$4$};
\end{tikzpicture}= O(\epsilon^{-1})
\]
is less divergent than other negative geometry at $L=4$. 
It will be interesting to see that directly from its negative geometry structure, which can be generalized to more complicated diagrams.
As inspired by this diagram, we can make a wild conjecture that even for higher loop negative geometry, any bipartite diagram with loops would diverge as $\epsilon^{-1}$ (or even $\epsilon^0$). Since when the number of loop momenta is odd, the integration of negative geometry diverges as $\epsilon^{-1}$ ~\cite{He:2023exb,Henn:2023pkc}. Therefore, the next non-trivial check will be at $L=6$, where we have already seen that the integration of $L=6$ ladder has the $\epsilon^{-2}$ divergence. If we are lucky enough that the conjecture was correct, we only need to consider the tree bipartite diagrams of negative geometries when computing the cusp anomalous dimension.

Although the negative geometry with an odd number of loop momenta do not contribute the cusp anomalous dimension, it is still interesting to integrate it at $L=5$. The first integrated $5$-loop negative geometry is the ladder $\tilde L_5$ eq.\eqref{5loopladder}, while the most complicated negative geometry would correspond to a bipartite graph with two loops 
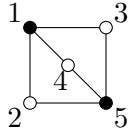
\[
\begin{tikzpicture}[baseline={([yshift=-.5ex]current bounding box.center)}]
\draw (-2,-0.5) -- (-1,-1.5);
\draw (-2,-1.5) -- (-1,-1.5) -- (-1,-0.5) -- (-2,-0.5) -- cycle;
\draw[fill=black] (-1,-1.5) circle[radius=.5ex];
\draw[fill=white] (-1,-0.5) circle[radius=.5ex];
\draw[fill=black] (-2,-0.5) circle[radius=.5ex];
\draw[fill=white] (-2,-1.5) circle[radius=.5ex];
\draw[fill=white] (-1.5,-1) circle[radius=.5ex];
\node at (-2.2,-0.3) {$1$};
\node at (-2.2,-1.7) {$2$};
\node at (-0.8,-0.3) {$3$};
\node at (-1.6,-1.2) {$4$};
\node at (-0.8,-1.7) {$5$};
\end{tikzpicture}
\]
with the integrand 
\[
\begin{aligned}
T_6&=
\frac{4}{c
s_1 t_2 t_3 t_4 s_5 D_{1,2} D_{1,3} D_{1,4} D_{2,5} D_{3,5} D_{4,5}}\bigl( -8 \epsilon_1 \epsilon_2\epsilon_3 \epsilon_4 \epsilon_5 N_{15}^s+ c \epsilon_2 \epsilon_3 \epsilon_4 P_a \nonumber\\
&+ c [ \epsilon_1 \epsilon_2 \epsilon_3 P_b + (\ell_1\leftrightarrow \ell_5)+ \epsilon_1 \epsilon_2 \epsilon_5 P_c + {\rm cyc} (\ell_2, \ell_3, \ell_4)]\nonumber\\
&  + c^2 [\epsilon_1 P_d + (\ell_1\leftrightarrow \ell_5)]+c^2 [\epsilon_2 P_e + {\rm cyc} (\ell_2, \ell_3, \ell_4)] \bigr),
\end{aligned}
\]
which is first computed in \cite{He:2022cup},
where 
\[
\begin{aligned}
P_a&:=-20 s_1 s_5+16 t_1 t_5+(N_{15}^s)^2, \quad P_b:=6 s_5 N_{14}^s,
P_c:=N_{15}^s N_{34}^t-4 N^{\rm cyc}_{1,3,5,4},\\
P_d&:=-s_5 (N_{12}^s N_{34}^t + {\rm cyc}(\ell_2, \ell_3, \ell_4))+[2\langle \ell_5 1 2\rangle^2 \langle \ell_1 1 2\rangle \langle \ell_2 3 4\rangle \langle \ell_3 3 4\rangle \langle \ell_4 3 4\rangle + {\rm cyc}(1,2,3,4)]\nonumber\\
&+ 2t_5 [\langle \ell_1 1 4\rangle (\langle \ell_2 1 4\rangle \langle \ell_3 2 3 \rangle \langle \ell_4 2 3 \rangle + {\rm cyc} (\ell_2, \ell_3, \ell_4))  + (14 \leftrightarrow 23)],\\
P_e&:=2 s_1 s_5 (N_{34}^t-N_{34}^s)-4 t_1 t_5 N_{34}^t-[s_5 (\langle \ell_1 1 2\rangle^2 
\langle \ell_3 3 4\rangle \langle \ell_4 3 4\rangle 
+ (12 \leftrightarrow 34)) +(\ell_1 \leftrightarrow \ell_5)]\nonumber\\
&+N_{15}^s (\langle \ell_1 1 4\rangle \langle \ell_5 1 4\rangle \langle \ell_3 2 3 \rangle \langle \ell_4 2 3\rangle +(14 \leftrightarrow 23)).
\end{aligned}
\]
It is important to look for a nice way to integrate negative geometries starting at $L=5$, it seems that both canonical differential equations \cite{Lagares:2024epo} and Mellin transformation used in this note will become not processable. The most natural way is directly integrating it in momentum twistor space where we still do not understand how to take the contour. Since we only need look for 3 contours for each 3D loop momentum, the ABJM theory will be a nice playground to test it.

For ladder diagrams we consider in this note, we provide a conjecture of recursion relation from $\tilde L_{2n}$ to $\tilde L_{2n+2}$, and we prove that they only contribute power of $\pi$ to the cusp anomalous dimension while there are other zeta values in their functions.
Since some derivation is not so 
rigorous, we need more checks for these conjectures. One possible check is to integrate the loop momenta in different orders, which gives the same leading divergence of $\epsilon^{-2}$. 

\section*{Acknowledgement} Zhenjie Li thanks Song He, Chia-Kai Kuo, Yichao Tang, Qinglin Yang,  Yao-Qi Zhang for stimulating discussions and collaborations on related projects, and thanks Bernhard Mistlberger for discussion on the numerical evaluation of 3D Feynman integrals.
We are grateful to Martin Lagares and Shun-Qing Zhang for sharing their results with us. This work is  supported in part by the U.S. Department of Energy under contract number DE-AC02-76SF00515. 

{\bf Note added}: While this work was in progress, we were made aware of an upcoming work~\cite{Lagares:2024epo}, where the finite function up to $L=4$ has been computed and the two results agree. 
\appendix

\section{Collection of useful 1-loop integrals}

We need two 1-loop integrals in this note
\begin{equation}\label{A1}
I_1=\int_{\ell} \frac{1}{(\ell\cdot 1)^{q_1}(\ell\cdot 2)^{q_2}(\ell \cdot 3)^{q_3}(\ell \cdot 4)^{q_4}(\ell\cdot \ell')^{q_5}}
\end{equation}
with $q_1+\cdots+q_5=3$ and 
\begin{equation}
I_2=\int_{\ell} \frac{N[\ell]}{(\ell\cdot 1)^{q_1}(\ell\cdot 2)^{q_2}(\ell \cdot 3)^{q_3}(\ell \cdot 4)^{q_4}(\ell\cdot \ell')^{q_5}}
\end{equation}
with $q_1+\cdots+q_5=4$,
where $N[\ell]$ is linear in dual momentum $\ell$ and $N[1]=N[3]=0$, for example, $N[\ell]$ could be $\epsilon_\ell$ or $(\ell\cdot 2)$ or $(\ell\cdot 4)$. 

Two special cases is scale triangle
$$
\int_{\ell} \frac{1}{(\ell\cdot a)(\ell\cdot b)(\ell \cdot c)}=\frac{\pi^{3/2}}{\sqrt{(a\cdot b)(b\cdot c)(c\cdot a)}}
$$
and the box integral with a numerator $N[\ell]$, 
$$
\int_{\ell} \frac{N[\ell]}{(\ell\cdot \ell_1)(\ell\cdot \ell_3)(\ell \cdot 1)(\ell\cdot 3)}=\pi^{3/2}\frac{N[\ell_1]/\sqrt{(\ell_1\cdot 1)(\ell_1\cdot 3)}+N[\ell_3]/\sqrt{(\ell_3\cdot 1)(\ell_3\cdot 3)}}{\sqrt{(1\cdot 3)(\ell_1\cdot \ell_3)}+\sqrt{(\ell_1\cdot 3)(\ell_3\cdot 1)}+\sqrt{(\ell_1\cdot 1)(\ell_3\cdot 3)}}.
$$

The most general cases can be done by introducing Feynman parameterization, 
\[
A=a_5\ell'+\sum_{i=1}^4a_iX_i,
\]
then for $I_1$
\begin{equation*}
\frac{1}{(\ell\cdot 1)^{q_1}(\ell\cdot 2)^{q_2}(\ell \cdot 3)^{q_3}(\ell \cdot 4)^{q_4}(\ell\cdot \ell')^{q_5}}=\int \frac{d^5a}{\text{GL}(1)}\frac{\Gamma(3)}{(\ell\cdot A)^3}\prod_i \frac{a_i^{q_i-1}}{\Gamma(q_i)}
\end{equation*}
and for $I_2$
\begin{equation*}
\frac{N]\ell]}{(\ell\cdot 1)^{q_1}(\ell\cdot 2)^{q_2}(\ell \cdot 3)^{q_3}(\ell \cdot 4)^{q_4}(\ell\cdot \ell')^{q_5}}=-\int \frac{d^5a}{\text{GL}(1)}\frac{\Gamma(3)N[\partial_{A}]}{(\ell\cdot A)^3}\prod_i \frac{a_i^{q_i-1}}{\Gamma(q_i)},
\end{equation*}
where the loop integral becomes trivial, and after integrating $\ell$ and Feynman parameters $a_1$, $a_2$ and $a_5$, we get that 
\begin{equation}
\begin{aligned}
I_1=\,&(1\cdot 3)^{-q_1-q_2-q_3+\frac{3}{2}} (2\cdot 4)^{-q_4}(\ell'\cdot 1)^{q_2+q_3-\frac{3}{2}} (\ell'\cdot 2)^{q_4-q_2} (\ell'\cdot 3)^{q_1+q_2-\frac{3}{2}}\\
&  \frac{\Gamma (-q_1-q_2+3/2)}{\pi ^{3/2} \Gamma (q_3) \Gamma (-q_1-q_2-q_3-q_4+3) \Gamma (q_4)}\\
&\int_{\mathbb R_+^2} da_3da_4\,a_3^{q_3-1} (a_3+1)^{-q_1} a_4^{q_4-1} (a_4+1)^{-q_2} (a_4 z'+a_3)^{q_1+q_2-\frac{3}{2}} 
\end{aligned}
\end{equation}
and 
\begin{equation}
\begin{aligned}
I_2=\,&(1\cdot 3)^{-q_1-q_2-q_3+\frac{5}{2}} (2\cdot 4)^{-q_4} (\ell'\cdot 1)^{q_2+q_3-\frac{5}{2}} (\ell'\cdot 2)^{q_4-q_2} (\ell'\cdot 3)^{q_1+q_2-\frac{5}{2}}\\
&\frac{\Gamma (-q_1-q_2+5/2) N[\ell']}{\pi ^{3/2} \Gamma (q_3) \Gamma (-q_1-q_2-q_3-q_4+4) \Gamma (q_4) }\\
&\int_{\mathbb R_+^2} da_3da_4\,a_3^{q_3-1} (a_3+1)^{-q_1} a_4^{q_4-1} (a_4+1)^{-q_2} (a_4 z'+a_3)^{q_1+q_2-\frac{5}{2}}.
\end{aligned}
\end{equation}
The $a_3$,$a_4$-integral can be further integrated if we introduce the Mellin transform of $(a_4z'+z_3)^k$
\[
(a_4z'+z_3)^k=\int_{\gamma}\frac{ds}{2\pi i} (z')^{-s}a_4^{-s}  a_3^{k+s}\frac{ \Gamma (s)\Gamma (-k-s)}{\Gamma (-k)},
\]
and we get that 
\[
\begin{aligned}
I_1=\,&(1\cdot 3)^{-q_1-q_2-q_3+\frac{3}{2}} (2\cdot 4)^{-q_4}(\ell'\cdot 1)^{q_2+q_3-\frac{3}{2}} (\ell'\cdot 2)^{q_4-q_2} (\ell'\cdot 3)^{q_1+q_2-\frac{3}{2}}\int_{\gamma}\frac{ds}{2\pi i} (z')^{-s}\\
&  \frac{\Gamma (s) \Gamma (-s-q_1-q_2+\frac{3}{2}) \Gamma (-s-q_2-q_3+\frac{3}{2}) \Gamma (s+q_1+q_2+q_3-\frac{3}{2}) \Gamma (s+q_2-q_4) \Gamma (q_4-s)}{\pi ^{3/2} \Gamma (q_1) \Gamma (q_2) \Gamma (q_3) \Gamma (-q_1-q_2-q_3-q_4+3) \Gamma (q_4)},
\end{aligned}
\]
and 
\[
\begin{aligned}
I_2=\,&(1\cdot 3)^{-q_1-q_2-q_3+\frac{5}{2}} (2\cdot 4)^{-q_4} (\ell'\cdot 1)^{q_2+q_3-\frac{5}{2}} (\ell'\cdot 2)^{q_4-q_2} (\ell'\cdot 3)^{q_1+q_2-\frac{5}{2}}N[\ell']\int_{\gamma}\frac{ds}{2\pi i} (z')^{-s}\\
&\frac{\Gamma (s) \Gamma (-s-q_1-q_2+\frac{5}{2}) \Gamma (-s-q_2-q_3+\frac{5}{2}) \Gamma (s+q_1+q_2+q_3-\frac{5}{2}) \Gamma (s+q_2-q_4) \Gamma (q_4-s)}{\pi ^{3/2} \Gamma (q_1) \Gamma (q_2) \Gamma (q_3) \Gamma (-q_1-q_2-q_3-q_4+4) \Gamma (q_4)}.
\end{aligned}
\]

\section{Function of 5-loop ladder}

Let $z=(1-x)^2/(4x)$, the 5-loop ladder is given by the integral
\[
\tilde L_5(z)\propto 
\int_{0<y<1} d\log\left(\frac{1-x y}{y-x}\right)\log \left(\frac{(x-1)^2 y}{x (y-1)^2}\right)\left(\log\biggl(\frac{(1-y)^2}{y}\biggr)^2+\frac{2 \pi ^2}{3}\right),
\]
and the result of this integral in generalized polylogarithms is
\[
\begin{aligned}
&\frac{2}{3} \pi ^2 G(x^{-1},0;1)-\frac{4}{3} \pi ^2 G(x^{-1},1;1)-\frac{2}{3} \pi ^2 G(x,0;1)+\frac{4}{3} \pi ^2 G(x,1;1)+6 G(x^{-1},0,0,0;1)-\\
&12 G(x^{-1},0,0,1;1)-12 G(x^{-1},0,1,0;1)+24 G(x^{-1},0,1,1;1)-12 G(x^{-1},1,0,0;1)+\\
&24 G(x^{-1},1,0,1;1)+24 G(x^{-1},1,1,0;1)-48 G(x^{-1},1,1,1;1)-6 G(x,0,0,0;1)+\\
&12 G(x,0,0,1;1)+12 G(x,0,1,0;1)-24 G(x,0,1,1;1)+12 G(x,1,0,0;1)-\\
&24 G(x,1,0,1;1)-24 G(x,1,1,0;1)+48 G(x,1,1,1;1)+\log (x+x^{-1}-2) \times\\&\biggl(2 G(x^{-1},0,0;1)-4 G(x^{-1},0,1;1)-4 G(x^{-1},1,0;1)+8 G(x^{-1},1,1;1)-2 G(x,0,0;1)+\\
&4 G(x,0,1;1)+4 G(x,1,0;1)-8 G(x,1,1;1)+\frac{2}{3} \pi ^2 \log (1-x)-\frac{2}{3} \pi ^2 \log (1-x^{-1})\biggr),
\end{aligned}
\]
and its symbol equals 
\[
\begin{aligned}
&16 [1-x,x,1-x,1-x]-8 [1-x,x,1-x,x]-8 [1-x,x,x,1-x]+\\
&4 [1-x,x,x,x]-8 [x,x,1-x,1-x]+4 [x,x,1-x,x]+4 [x,x,x,1-x]-2 [x,x,x,x],
\end{aligned}
\]
where we use $[a,b,c,d]$ to represent $a\otimes b\otimes c\otimes d$.

\bibliographystyle{utphys}
\bibliography{refs}

\end{document}